# Supernova Simulations and Strategies
# For the Dark Energy Survey

J. P. Bernstein[1], R. Kessler[2,3], S. Kuhlmann[1],
R. Biswas[1], E. Kovacs[1], G. Aldering[4], I. Crane[1,5], C. B. D'Andrea[6],
D. A. Finley[7], J. A. Frieman[2,3,7], T. Hufford[1], M. J. Jarvis[8,9], A. G. Kim[4],
J. Marriner[7], P. Mukherjee[10], R. C. Nichol[6], P. Nugent[4], D. Parkinson[10],
R. R. R. Reis[7,11], M. Sako[12], H. Spinka[1], M. Sullivan[13]

## ABSTRACT

We present an analysis of supernova light curves simulated for the upcoming Dark Energy Survey (DES) supernova search. The simulations employ a code suite that generates and fits realistic light curves in order to obtain distance modulus/redshift pairs that are passed to a cosmology fitter. We investigated several different survey strategies including field selection, supernova selection biases, and photometric redshift measurements. Using the results of this study, we chose a 30 square degree search area in the *griz* filter set. We forecast 1) that this survey will provide a homogeneous sample of up to 4000 Type Ia supernovae in the redshift range 0.05<z<1.2, and 2) that the increased red efficiency of the DES camera will significantly improve high-redshift color measurements. The redshift of each supernova with an identified host galaxy will be obtained from spectroscopic observations of the host. A supernova spectrum will be obtained for a subset of the sample, which will be utilized for control studies. In addition, we have investigated the use of combined photometric redshifts taking into account data from both the host and supernova. We have investigated and estimated the likely contamination from core-collapse supernovae based on photometric identification, and have found that a Type Ia supernova sample purity of up to 98% is obtainable given specific assumptions. Furthermore, we present systematic uncertainties due to sample purity, photometric calibration, dust extinction priors, filter-centroid shifts, and inter-calibration. We conclude by estimating the uncertainty on the cosmological parameters that will be measured from the DES supernova data.

*Subject headings:* supernovae – cosmology: simulations

[1]Argonne National Laboratory, 9700 South Cass Avenue, Lemont, IL 60439, USA

[2]Kavli Institute for Cosmological Physics, The University of Chicago, 5640 South Ellis Avenue Chicago, IL 60637, USA

[3]Department of Astronomy and Astrophysics, The University of Chicago, 5640 South Ellis Avenue Chicago, IL 60637, USA

[4]E. O. Lawrence Berkeley National Laboratory, 1 Cyclotron Rd., Berkeley, CA 94720, USA

[5]Department of Physics, University of Illinois at Urbana-Champaign, 1110 West Green Street, Urbana, IL 61801-3080 USA

[6]Institute of Cosmology and Gravitation, University of Portsmouth, Dennis Sciama Building, Burnaby Road, Portsmouth PO1 3FX, UK

[7]Center for Particle Astrophysics, Fermi National Accelerator Laboratory, P.O. Box 500, Batavia, IL 60510, USA

[8]Centre for Astrophysics, Science & Technology Research Institute, University of Hertfordshire, Hatfield, Herts, AL10 9AB, UK

[9]Physics Department, University of the Western Cape, Cape Town, 7535, South Africa

[10]Department of Physics and Astronomy, Pevensey 2 Building University of Sussex, Falmer Brighton BN1 9QH, UK

[11]Now at: Instituto de Física, Universidade Federal do Rio de Janeiro C. P. 68528, CEP 21941-972, Rio de Janeiro, RJ, Brazil

[12]Department of Physics and Astronomy, University of

Contents

Pennsylvania, 203 South 33rd Street, Philadelphia, PA 19104, USA
[13]Department of Physics, Denys Wilkinson Building, Oxford University, Keble Road, Oxford, OX1 3RH, UK

1. Introduction

In the late 1990's, observations of distant Type Ia supernovae (SNIa) provided the convincing evidence for the acceleration of cosmic expansion (Riess et al. 1998; Perlmutter et al. 1999). Dedicated supernova (SN) surveys covering cosmologically relevant redshifts, such as the ESSENCE Supernova Survey (Miknaitis et al. 2007; Foley et al. 2009), Supernova Legacy Survey (SNLS, Astier et al. 2006; Conley et al. 2011), Sloan Digital Sky Survey-II Supernova Survey (SDSS, Frieman et al. 2008b; Sako et al. 2011), Carnegie Supernova Project (Hamuy et al. 2006; Stritzinger et al. 2011), Stockholm VIMOS Supernova Survey II (Melinder et al. 2011), and Hubble Space Telescope searches (e.g., Strolger et al. 2004; Dawson et al. 2009; Amanullah et al. 2010), have substantially improved the quantity and quality of SNIa data in the last decade. A previously unknown energy-density component known as dark energy is the most common explanation for cosmic acceleration (for a review, see Frieman et al. 2008a; Weinberg et al. 2012). The recent SN data, in combination with measurements of the cosmic microwave background (CMB) anisotropy and baryon acoustic oscillations (BAO), have confirmed and constrained accelerated expansion in terms of the the relative dark energy density ($\Omega_{DE}$) and equation of state parameter ($w \equiv p_{DE}/\rho_{DE}$, where $p_{DE}$ and $\rho_{DE}$ are the pressure and density of dark energy, respectively). The next generation of cosmological surveys is designed to improve the measurement of $w$, and constrain its variation with time, from observations of the most powerful probes of dark energy as suggested by the Dark Energy Task

Force (Albrecht et al. 2006): SNe, BAO, weak lensing, and galaxy clusters.

Future SN surveys face common issues, including the number and position of fields, filters, exposure times, cadence, and spectroscopic and photometric redshifts. Each study must optimize telescope allocations to return the best cosmological constraints. The simulation analysis presented in the paper is for the Dark Energy Survey[14] (DES), which expects to see first-light in 2012. The DES will carry out a deep optical and near-infrared survey of 5000 square degrees of the South Galactic Cap (see Fig. 1) using a new 3 deg$^2$ Charge Coupled Device (CCD) camera (the Dark Energy Camera, or "DECam," Flaugher et al. 2010) to be mounted on the Blanco 4-meter telescope at the Cerro Tololo Inter-American Observatory (CTIO). The DES SN component will use approximately 10% of the total survey time during photometric conditions and make maximal use of the non-photometric time, for a total SN survey of ~1300 hours. The DECam focal plane detectors (Estrada et al. 2010) are thick CCDs from Lawrence Berkeley National Laboratory (LBNL), which are characterized by much improved red sensitivity relative to conventional CCDs (see Fig. 2, as well as Holland 2002; Groom et al. 2006; Diehl et al. 2008). This will allow for deeper measurements in the redder bands, which is of particular importance for high-redshift SNe. This effect is shown in Fig. 3, which plots simulated scatter in the `SALT2` (Guy et al. 2007) SN color parameter (see §5.2) for the SDSS, SNLS, and DES. Note, in particular, the superior high-redshift color measurements in the DES deep fields (see §3.1). Details of the simulation method can be found in §2. The implementation for the DES, e.g., an exposure time of approximately an hour in the SDSS-like $z$ passband per field per observation, is discussed in §3.

An accurate redshift determination (to ~0.5%) is necessary to place a SN on the Hubble diagram. This can be obtained by taking a spectrum of the SN itself or of its host galaxy. A spectrum of the SN has the added advantage of providing a definitive confirmation of the SN type, and allowing for studies of systematic variations, but is more difficult to obtain. Follow-up spectroscopy of the host galaxy can be done at a later date, taking advan-tage of multi-object spectrograph capabilities to obtain many spectra at once. Photometric redshifts can also be obtained using deep co-added photometry of the host galaxy, but the redshift accuracy is degraded, reducing the usefulness of the SN for cosmological measurements. The existing SNIa samples from previous surveys include a subset of SNe with measured spectra consisting of ~1000 SNIa spread out over many surveys (Sullivan et al. 2011; Amanullah et al. 2010, and references therein), and the remainder includes many more SNe with host spectra or host and/or SN photometric redshifts. The usefulness of the current photometric samples depends on the fraction of host galaxies that will be followed-up, a number which is uncertain. The DES will identify up to ~4000 high-quality SNIa, and plans a follow-up program to acquire SN spectra near peak for up to 20% of this sample and host galaxy spectra for the majority of the remainder. For SNe that do not have a follow-up or host galaxy spectrum taken, a deep co-add of images (>70 hours per season) will be used to determine the host photometric redshift. This host redshift will be further utilized as a prior for a combined SN photometric redshift fit.

---

[14] http://www.darkenergysurvey.org

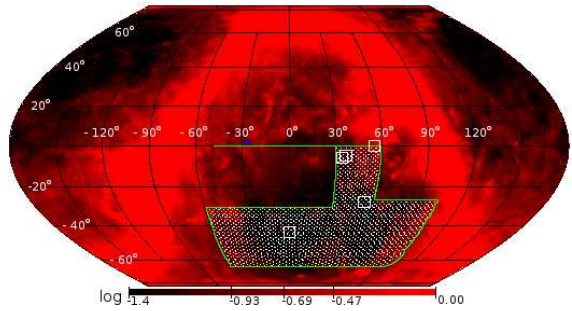

Fig. 1.—: The DES footprint. The white squares indicate the locations of our current choice of five SN fields (see §3.1). For the survey strategies considered in this analysis with additional shallow fields, those fields are placed next to these five fields. The size of the squares as shown is much larger than the 3 deg$^2$ field of view of DECam in order to make them easier to see in this Figure. The scale shows the log of $r$-band (as defined in §3.1) Galactic extinction in magnitudes.

In order to aid in the design of the DES SN search, we simulate expected DES SN observations. We use the parametric `SNANA` code suite

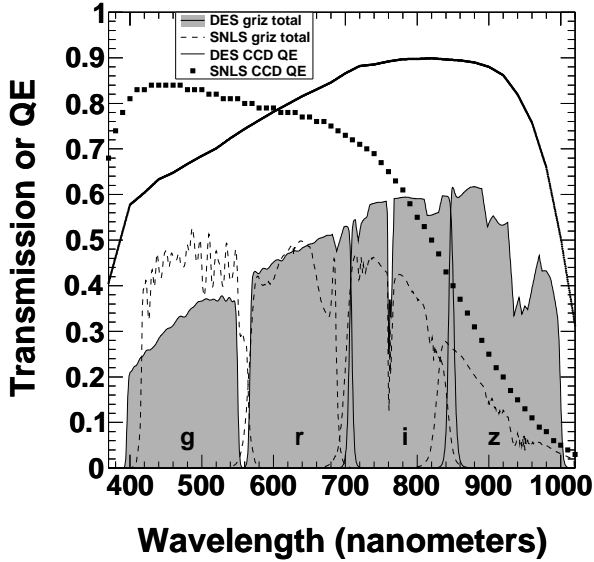

Fig. 2.—: Comparison of the SNLS (Regnault et al. 2009) and DECam total transmission (H. Lin, private communication, 2011) for an airmass of 1.3. Also shown is the CCD quantum efficiency (QE). The total transmission includes the effects of QE, the atmosphere, and the optical systems of the relevant cameras. Note the increased DES sensitivity at redder wavelengths. The DECam transmission is based on measurements of the full-size filters, which was not available during the simulations performed for this analysis. The assumed transmission in this paper is about 10% smaller than the measured values.

(Kessler et al. 2009b) that generates SN light curves using realistic models and takes into account, e.g., seeing conditions, Galactic extinction, and CCD noise. In this work, we use the optical ($\lambda < 1$ $\mu$m) `MLCS2k2` (Jha et al. 2007) and `SALT2` (Guy et al. 2007) models and the optical+infrared `SNooPy` model (Burns et al. 2011). We chose to employ the `MLCS2k2` model because the inclusion of the straightforward reddening parametrization from Cardelli et al. (1989) makes it easier to assess systematic errors simply by varying the parameters. In contrast, the parametrization in `SALT2` is more complex which complicates the systematics studies. We further employ a light curve fitter, based on `MLCS2k2` and `SALT2` models, to obtain a prediction of the measured distance modulus, $\mu$, for each SN. Measured redshifts are expected to come from a combination of spectra and photometric redshifts from the host galaxy and SN, and results are compared for these different scenarios.

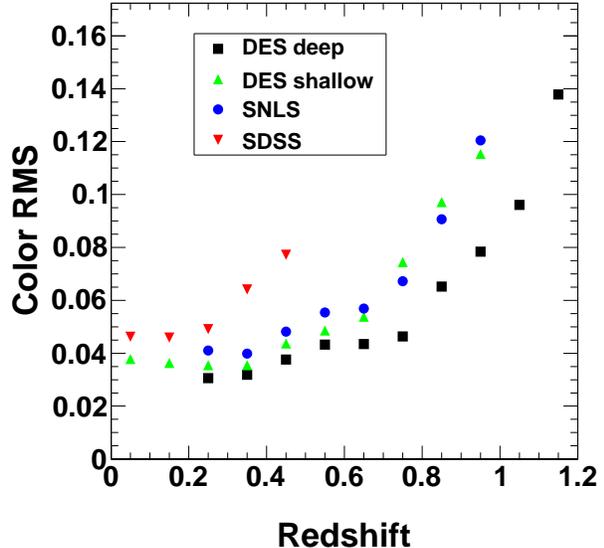

Fig. 3.—: Simulation of the scatter in the `SALT2` color parameter for the SDSS, SNLS, and DES supernova samples highlighting the red advantage of the DES. The simulation method and DES implementation are discussed in §2 and §3, respectively.

The outline of the paper is as follows. We present our method of SN light curve simulation in §2. We discuss the DES options and present example simulations in §3. Redshift determinations, both spectroscopic and photometric, are discussed in §4. Analysis options are presented in §5. A study of Type Ia sample purity is presented in §6. SN colors and dust extinction are discussed in §7, and projected cosmology constraints are presented in §8. Finally, we summarize and discuss our results in §9.

## 2. Supernova light curve simulation

In this section, we present our SN light curve simulations in greater technical detail. We discuss general properties of `SNANA` in §2.1 and introduce our application to the DES in §2.2.

## 2.1. SNANA

We employ the SNANA package (Kessler et al. 2009b) to simulate and fit Type Ia and Type Ibc/II SN light curves. We emphasize that while we are using SNANA to investigate the capabilities of the DES, it was originally developed for and utilized for the analysis of observational SDSS SN data (Kessler et al. 2009a), was used by the Large Synoptic Survey Telescope (LSST) collaboration to forecast SN observations (Abell et al. 2009), and can be applied to any survey in general. Using the simulation requires a survey-specific library that includes the survey characteristics, e.g., filters, observing cadence, seeing conditions, zeropoints, and CCD characteristics.

For a rest-frame SN light curve model, such as MLCS2k2, the basic simulation steps are as follows:

1. pick a sky position, redshift from observed SN rate distributions, and sequence of observer and rest-frame observation times;

2. pick SN luminosity ($\Delta$) and $V$-band host-extinction ($A_V$, the amount of dust extinction in magnitudes from Cardelli et al. 1989) parameters randomly drawn from their distributions;

3. generate a rest-frame light curve from the SN light curve model: e.g., magnitudes in the $U$, $B$, $V$, $R$, & $I$ filters (Bessell 1990) versus time;

4. add host-galaxy extinction to each rest-frame magnitude using $A_V$ (from Step 2 above) and the CCM dust model from Cardelli et al. (1989);

5. add K-corrections (Nugent et al. 2002) to transform $UBVRI$ to observer-frame filters[15];

6. add Galactic (Milky Way) extinction using data from Schlegel et al. (1998);

7. use survey zeropoints to translate above-atmosphere magnitudes into observed flux in CCD counts;

8. compute noise from the sky level, point spread function (PSF), CCD readout noise (negligible for the DES), and signal Poisson statistics;

9. In addition to steps 4 and 5 above, apply an ad-hoc Gaussian smearing model of intrinsic SN color variations to obtain Hubble residuals that match observations.

We make use of three light-curve models that are integrated into SNANA to simulate and fit SN light curves: MLCS2k2, SALT2, and SNooPy. Note that SNANA uses MINUIT (James 1994) for minimization. The MLCS2k2 model is improved relative to the Jha et al. (2007) code (see §5.1 and Appendix B of Kessler et al. 2009a), e.g., it fits in flux instead of magnitudes and includes simulated efficiency in the prior. A key difference between MLCS2k2 and SALT2 is that the former fits for a distance for each SN while the latter does not. The SALT2 light curve model in SNANA is accompanied by a separate program called SALT2mu (Marriner et al. 2011) that is used to determine a distance for each SN so that the MLCS2k2 and SALT2 fit results can be treated in the same way (see §5.2 for additional information).

## 2.2. Simulation inputs

Construction of a survey-specific library as input to the SN simulation is crucial to obtaining realistic simulated light curves. For each DES SN observing field, this library includes information about the survey cadence, filters, CCD gain and noise, PSF, sky background level, and zeropoints and their fluctuations. The zeropoint encodes exposure time, atmospheric transmission, and telescope efficiency and aperture. These quantities vary with each exposure and so, for the DES study, we created a program which uses, among other things, the CTIO weather histories, ESSENCE zeropoint and PSF data, time gaps due to Blanco community use, and Moon brightness to estimate the parameters for the DES simulation library. Table 1 shows example entries in this library, and we now discuss the details of their creation.

The SN component of the DES is limited to about 10% of the total survey photometric time. In all cases, after a certain period of time (expected to be ∼8 days), if a SN field has not been observed it becomes the top observational priority of the survey even under photometric conditions. There are two main options being considered for the decision procedure to observe in shorter intervals than 8 days: 1) make maximal use of non-photometric time based on an infrared cloud camera (RASICAM, Lewis et al. 2010), or 2) decide based on the measured seeing, giving the non-SN DES components the best seeing for weak lensing and other science, and switch to the SN fields if the PSF is $\gtrsim 1''$. The final DES decision tree will probably be a combination of these two. In this analysis we have simulated option #1.

---
[15]K-corrections are needed in both the simulator and fitter, and are applied using a technique very similar to that in Jha et al. (2007).

| MJD/Filter | PSF (pixels) | $\sigma_{\rm sky}$ ($e^-$) | Zpt (mag) |
|---|---|---|---|
| 55881.191/$g$ | 2.26 | 80 | 33.0 |
| 55881.199/$r$ | 2.16 | 151 | 34.5 |
| 55881.215/$i$ | 2.05 | 257 | 34.7 |
| 55881.238/$z$ | 1.79 | 651 | 35.6 |
| 55884.312/$g$ | 2.58 | 143 | 32.7 |
| 55884.328/$r$ | 2.62 | 220 | 34.3 |
| 55884.344/$i$ | 2.35 | 390 | 34.4 |
| 55885.188/$z$ | 2.83 | 764 | 35.7 |

Table 1:: Example DES SN simulation inputs for a 4-day excerpt from a single, 6-month season where "$\sigma_{\rm sky}$" is the sky noise in photoelectrons and "Zpt" is the zeropoint in magnitudes. Additional inputs that are needed, but not shown in this Table, are the RA & DEC of the field, CCD gain and noise, pixel size, and the contribution to the zeropoint due to fluctuations.

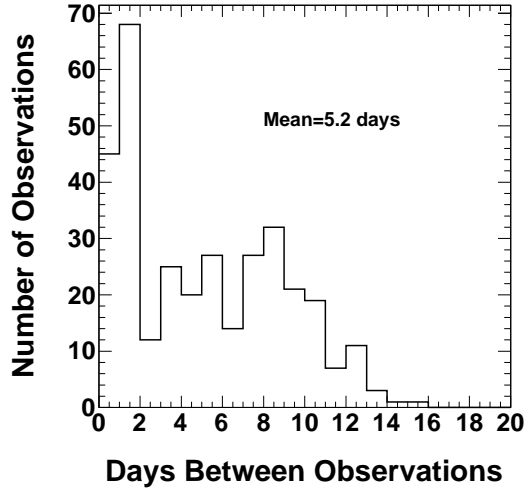

Fig. 4.—: The DES 5-field hybrid strategy (see Tab. 3) forecast for the distribution of the temporal gaps between observations during a typical DES SN season for the two deep fields only. The histogram entries are for all of the DES SN filters combined, e.g., a gap of 6 days in observations in any one of the filters increments the count of 6-day gaps. The gaps in the 10-field hybrid strategy are very similar.

The separation of non-photometric and photometric time in the generation of the simulation library is accomplished by incorporating weather history maintained at CTIO for more than twenty years. This SN survey strategy leads to a two-component cadence: a peak in the number of observations at very short cadence due to several-day periods of non-photometric time when the SN fields dominate the observing time, and a broad secondary maximum around 8 days when photometric time is used (see Fig. 4). Our SN observing also requires an airmass less than 2.0. This, combined with DES half-nights in January and February and long periods of photometric conditions, can lead to cadences longer than 8 days.

The other critical components of the simulation input library are the PSF, sky background, and zeropoints. Usually, in this type of study, one takes averages of these quantities. In our case, we have used ESSENCE data to provide variations of the PSF and zeropoints at CTIO for each observation, as well as SDSS data for the dependence of sky background on relative Moon position. The measured PSF variation of ESSENCE is input directly into the simulation library after correcting for the wavelengths of the filter centroids and airmasses for the mock DES observation. The resulting PSF distribution is shown in Fig. 5. Recent improvements to the telescope and its environment, along with the optical design and mechanical (hexapod) control of DECam, are expected to deliver improved image quality compared to these data. The choice of PSF distribution is conservative for option #1 and consistent with option #2.

Another key input to the simulation is the rate of SNIa explosions in the Universe as a function of redshift (see §6.1 for a discussion of the input rate of core-collapse SNe). The total number of SNe that the DES will observe is clearly directly sensi-

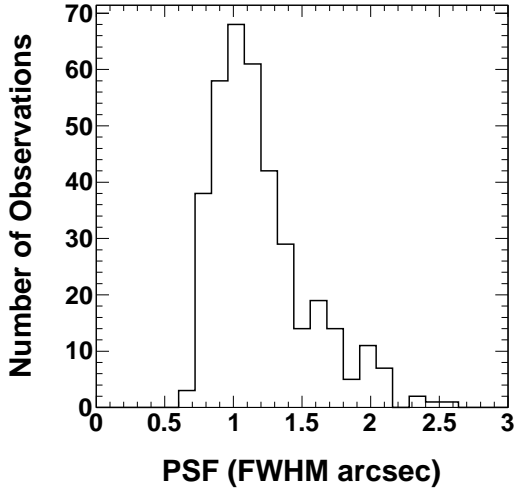 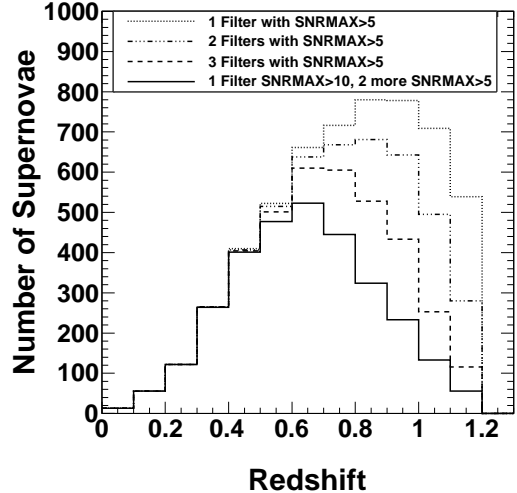

Fig. 5.—: The DES 5-field hybrid strategy (see Tab. 3) forecast for the number of observations versus the input PSF for a typical DES SN season. The histogram entries are for all of the DES SN filters combined, e.g., an observation with a PSF value of 1.0 in any one of the filters increments the count of 1.0 PSF observations. The PSF distribution is very similar for the other survey strategies.

Fig. 6.—: Type Ia supernova redshift distributions for the DES 5-field hybrid strategy (see Tab. 3) for the various SNRMAX cuts indicated in the legend. The total number of simulated SNe passing each set of cuts, from top to bottom, is 5571, 4783, 3906, and 3047.

tive to that rate. The default SNIa rate we employ in `SNANA` is the power law from Dilday et al. (2008):

$$R_{\rm SNIa} \equiv \text{SNIa rate} = \alpha_{\rm Ia} \times (1+z)^{\beta_{\rm Ia}} \qquad (1)$$

where $\alpha_{Ia} = (2.6+0.6-0.5)\times 10^{-5}$ SNe $h_{70}^3$ Mpc$^{-3}$ yr$^{-1}$, $h_{70} = H_0/(70$ km s$^{-1}$ Mpc$^{-1})$, where $H_0$ is the present value of the Hubble parameter[16], and $\beta_{Ia} = 1.5 \pm 0.6$. In addition, Dilday et al. (2008) further found the correlation coefficient between $\alpha_{Ia}$ and $\beta_{Ia}$ to be $-0.80$. Extrapolating this rate to redshifts greater than 1 is highly uncertain.

## 3. Survey strategy options and example simulations

Simulation of the current DES observing strategy leads to a total exposure time for the SN search of ∼1300 hrs, over 900 hrs of which occur during non-photometric conditions. On a given night, prioritization of observations in each of the *griz* filters will be made based on descending time since the previous observation. There is an automatic 8-day trigger if no photometric observation has been performed for a given filter. For the simulations presented in this section, we employed the `MLCS2k2` model as the basis for generating and fitting SN light curves over the redshift range of 0.0<z<1.2. The free parameters of the model are the time of maximum light in the B-band ($t_o$), the distance modulus ($\mu$), the luminosity parameter ($\Delta$), and the extinction in magnitudes by dust in the host galaxy (parametrized by $A_V$ and $R_V$ from Cardelli et al. 1989). In this section, $A_V$ and $\Delta$ were constrained to a range of 0.0 to 2.0 and $-0.4$ to 1.80, respectively, and $R_V$ was fixed to 2.18 (Kessler et al. 2009a). Parameter variations and comparisons with simulations using the `SALT2` model will be presented in later sections.

---

[16] We use $H_0 = 65$ km s$^{-1}$ Mpc$^{-1}$ in our `MLCS2k2` simulations to match the training value, and 70 km s$^{-1}$ Mpc$^{-1}$ in our `SALT2` simulations. These values of the Hubble parameter are also used to determine the simulated distance modulus in a flat $\Lambda CDM$ model with $\Lambda = 0.73$. We do not attempt to model rate differences due to host-galaxy type.

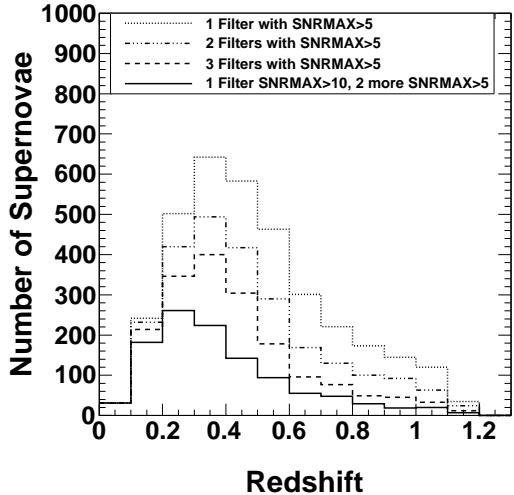

Fig. 7.—: Core-collapse supernova redshift distributions for the DES 5-field hybrid strategy (see Tab. 3) for the various SNRMAX cuts indicated in the legend. The total number of simulated SNe passing each set of cuts, from top to bottom, is 3458, 2462, 1785, and 1112.

### 3.1. Fields, filters, and selection cuts

The choice of the DES SN fields is driven by four primary considerations:

- visibility from CTIO,
- visibility from Northern-hemisphere, 8-meter-class telescopes for SN follow-up spectroscopy,
- past observation history as it pertains to the use of pre-existing galaxy catalogs and calibration,
- overlap with the survey area for the Visible & Infrared Survey Telescope for Astronomy (VISTA, Emerson et al. 2004, see §7.2).

Based on these criteria, we have tentatively chosen the five fields in Tab. 2 and Fig. 1. In this paper, we consider five SN survey strategies (see Tab. 3). For the 10-field hybrid strategy, the fields are the two deep fields and 3 shallow fields from the 5-field hybrid plus 5 additional shallow fields clustered around the *Chandra* Deep Field South field. In later sections we will compare in detail the results of these surveys, including projected constraints on cosmology.

| Field (3 deg$^2$ area) | Pointing RA&Dec (deg., J2000) |
|---|---|
| *Chandra* Deep Field S. | $52.5°, -27.5°$ |
| XMM-LSS | $34.5°, -5.5°$ |
| SDSS Stripe 82 | $55.0°, 0.0°$ |
| SNLS D1/Virmos VLT | $36.75°, -4.5°$ |
| ELAIS S1 | $0.5°, -43.0°$ |

Table 2:: Likely Dark Energy Survey supernova fields. Note that not all of these fields satisfy all of the field choice optimization criteria discussed in the text; e.g., ELAIS S1 is not visible from Northern-hemisphere, 8-meter-class telescopes, but matches the other criteria well.

| Survey strategy | # deep fields | # shallow fields | Area (deg$^2$) |
|---|---|---|---|
| ultra-deep | 1 | 0 | 3 |
| deep | 3 | 0 | 9 |
| shallow | 0 | 9 | 27 |
| 5-field hybrid | 2 | 3 | 15 |
| 10-field hybrid | 2 | 8 | 30 |

Table 3:: Dark Energy Survey supernova strategies considered in this paper where each SN field has an area equal to the DECam 3 deg$^2$ field of view. Note that the difference between deep and shallow fields is exposure time, not area.

We used SNANA to explore the choice of filters and exposure times, and the resulting effects on survey cadence. We evaluated the effect of the $griz$ and $grizY$ filter sets on DES SN observations. Figure 2 shows the chosen DES SN filters along with the DES CCD quantum efficiency. In this paper, we have selected five SN search strategies that span the range from ultra-deep and narrow to wide and shallow, including hybrid mixtures of the two. Table 4 shows the filter exposure times for the deep fields for the deep and 10-field hybrid strategies and the shallow fields for the 10-field hybrid strategy for the $griz$ filter set (see the discussion about $Y$-band at the end of this section). Table 5 shows the limiting magnitudes in each filter for the 10-field hybrid survey. For all the survey strategies considered, the deep fields have the exposure times listed in the second column of Tab. 4. For the shallow survey considered in this paper, as well as for the shallow fields in the 5-field hybrid strategy, each of the fields has one third of the total exposure time per field of a deep field.

| Filter | Deep exp. time (s) | Lim. mag. | Shallow exp. time (s) | Lim. mag. |
|---|---|---|---|---|
| $g$ | 300 | 25.2 | 175 | 24.9 |
| $r$ | 1200 | 25.4 | 50 | 23.7 |
| $i$ | 1800 | 25.1 | 200 | 23.9 |
| $z$ | 4000 | 24.9 | 500 | 23.8 |

Table 4:: Filter exposure times and limiting magnitudes for the 10-field hybrid strategy. The deep and shallow times were chosen to roughly equalize signal-to-noise at high redshift and near a redshift of z=0.5, respectively (see Fig. 8). Limiting magnitudes are for point sources detected at $5\sigma$ using a single filter observation.

| Filter | Deep Fields | Shallow Fields |
|---|---|---|
| $g$ | 27.1 | 26.8 |
| $r$ | 27.3 | 25.6 |
| $i$ | 27.0 | 25.9 |
| $z$ | 26.8 | 25.7 |

Table 5:: Limiting magnitudes for point sources detected at $5\sigma$ in the DES 10-field hybrid survey, using a 1-season co-add and assuming 35 filter observations per season. The limiting magnitudes for a 5-season co-add are $\sim 0.85$ magnitudes deeper.

We define "epoch" to be an observation in a single filter on a given date (with no requirement on a source detection). In order to produce simulated sets of DES SN light curves that realistically represent the quality needed for the determination of cosmological parameters, we defined selection cuts that each simulated light curve must individually satisfy (see Tab. 6). The selection criteria ensure that a DES SN light curve used for analysis is well-sampled, with measurements both when the light curve is rising and falling, and of sufficient quality to allow for a robust distance determination, which is essential for constraining cosmology. However, these cuts are relatively inefficient for SNIa retention at higher DES redshifts; studies of the use of looser cuts in conjunction with photometric SN typing methods are ongoing. The effects of different cuts on the maximum signal-to-noise in a given pass-band (SNRMAX) on simulated Type Ia and simulated core collapse samples (described in more detail later) are shown in Fig. 6 and Fig. 7 respectively. The tightest cuts shown, which are our defaults in this paper, produce the best sample purity at the expense of lower SNIa efficiency.

| Selection cuts for DES SNe |
|---|
| 1. At least 5 total epochs above a signal-to-noise threshold of 0.01; |
| 2. At least one epoch before and at least one 10 rest-frame days after the $B$-band peak; |
| 3. At least one filter measurement with a SNRMAX above 10; |
| 4. At least two additional filter measurements with a SNRMAX above 5. |

Table 6:: Selection cuts that each simulated light curve must individually satisfy in order to ensure realistic simulations of the DES SN capabilities. Note that epochs that are included in the light curve fit are between a rest-frame phase of $-15$ and $+80$ days.

Fig. 8 shows example multi-band SNRMAX values for simulated DES SN light curves subject to the cuts described above assuming the 10-field hybrid strategy. Note how the $g$-band measurements have significantly reduced SNRMAX beyond a redshift of z$\sim$0.5 and are absent beyond z$\sim$0.8 due to the flux being redshifted out of the wavelength range of the light curve model.

Our investigation of a $grizY$ survey option showed that the $Y$-band SNRMAX barely reaches above 5 even when half of the deep $z$-band exposure time is devoted to it, and that the $Y$-band drops below SNRMAX of 5 at a redshift of $\sim 0.7$. Thus, we elected not to use the $Y$ filter for DES SN observations. Note, however, that the planned DES overlap with the VIDEO Survey will provide for $Y$-band and $J$-band light curves for a few percent of the DES SNe (see §7.2).

## 3.2. Light curves and SN statistics

Fig. 9 shows example DES light curves at redshifts of 0.25, 0.50, 0.74, and 1.07. Particularly noteworthy is that the flux errors projected for DES SN observations are very small at lower redshifts and remain reasonable even beyond a redshift of z=1. The fact that the *g*-band is absent for the z=0.74 and z=1.07 light curves highlights why high-redshift SNe only have 3 pass-bands for *griz* surveys.

A key to planning a cosmological SN search is the trade-off between survey area and depth. For the DES SN search, a motivation for deep observations is the advantage of the DECam red sensitivity, while a wide survey area is desirable because it returns a greater number of SNIa at a given signal-to-noise. In other words, the observing strategy should be both wide, to maximize SN statistics, and deep, to provide for a longer lever arm. Fig. 10 shows the SNIa redshift distribution for the deep, shallow, and two hybrid survey strategies. We also considered an ultra-deep strategy (3 deg$^2$). We found that the ultra-deep strategy delivers only a marginal improvement in SNIa statistics beyond a redshift of z=1 relative to the 10-field hybrid strategy, for example, while the latter results in a factor of 2.8 more SNIa overall. In particular, we found that the 10-field hybrid has 42% more SNIa in the redshift range of 0.6-1.0 relative to the ultra-deep strategy. In addition, the ultra-deep survey produces statistics inferior to the deep survey. Thus, the ultra-deep strategy is withdrawn from consideration, while the deep strategy is carried throughout this paper. Figure 10 also shows that the deep and shallow surveys exhibit a significant decrease in the number of SNe at low- and high-redshifts, respectively, relative to the two hybrid surveys. The hybrid surveys also retain a significant fraction of the low- and high-redshift SNe found in the shallow and deep surveys while avoiding a significant fraction of the selection bias of the shallow survey (see §5.1). The redshift distributions for the hybrid surveys including the deep and shallow components are shown in Fig. 11. The 10-field hybrid strategy is preferred on the grounds of maximizing SN statistics in the intermediate redshift regime.

In order to explore the sensitivity of the redshift distribution to the rate of SNIa, we performed simulations including the $\alpha_{Ia}$ and $\beta_{Ia}$ variations according to the uncertainties given by Eqn. 1. Since Dilday et al. (2008) found the correlation coefficient between $\alpha_{Ia}$ and $\beta_{Ia}$ to be $-0.80$, we ran simulations assuming the parameters are 100% anti-correlated. We found that the projected number of DES SNIa would change by approximately 7% given such a rate variation.

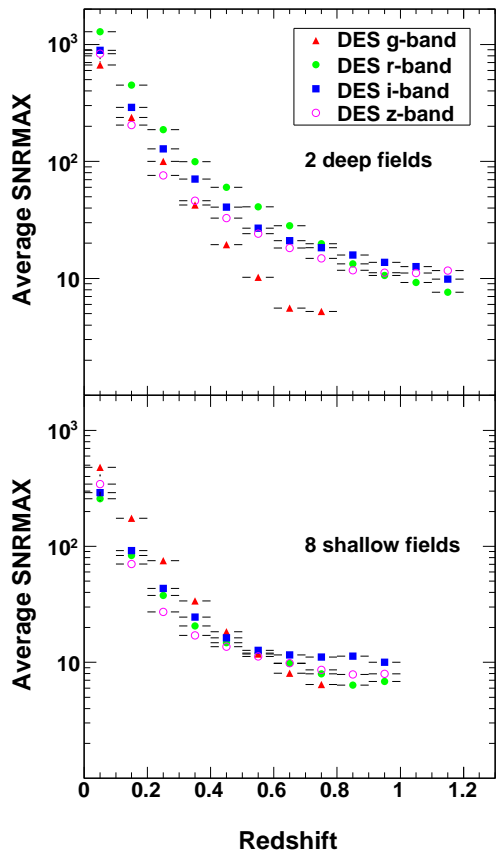

Fig. 8.—: Average maximum signal-to-noise for SNIa in a given pass-band (SNRMAX) for the 10-field hybrid strategy as a function of redshift in the DES $g$-, $r$-, $i$-, and $z$-bands. Note that at higher redshifts, the points are effected by the selection criteria. The upper and lower panels show the result for the deep and shallow fields, respectively.

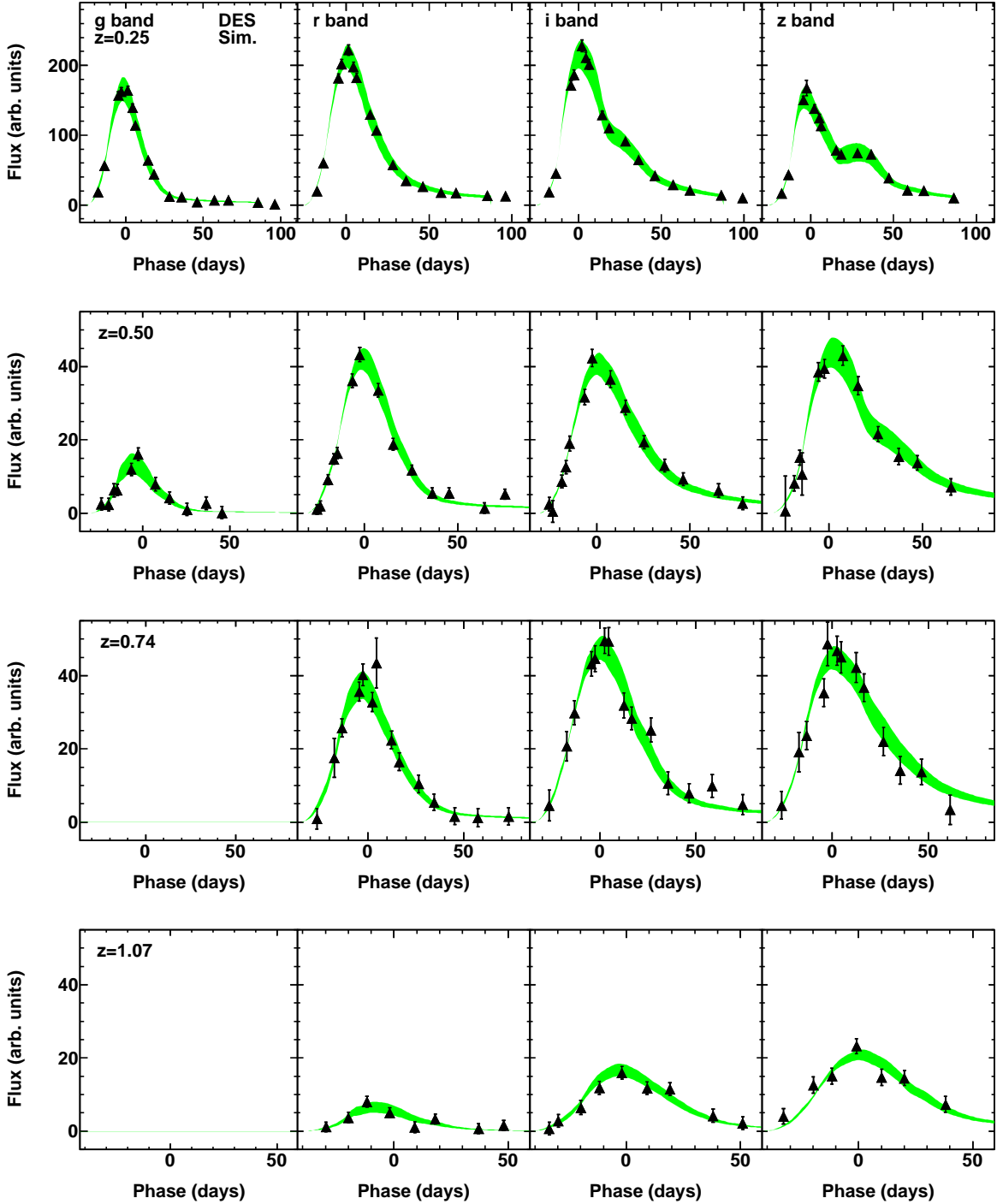

Fig. 9.—: From *top* to *bottom*: simulated DES light curves for the deep component of the 5-field hybrid strategy at redshifts of z=0.25, 0.50, 0.74, and 1.07, respectively. The points are MLCS2k2 simulated data, the center of the band is the MLCS2k2 fit, and the width of the band gives the fit error. Note the flux accuracy and progressive reduction in *g*-band flux until it drops out entirely due being redshifted out of the wavelength range of the light curve model.

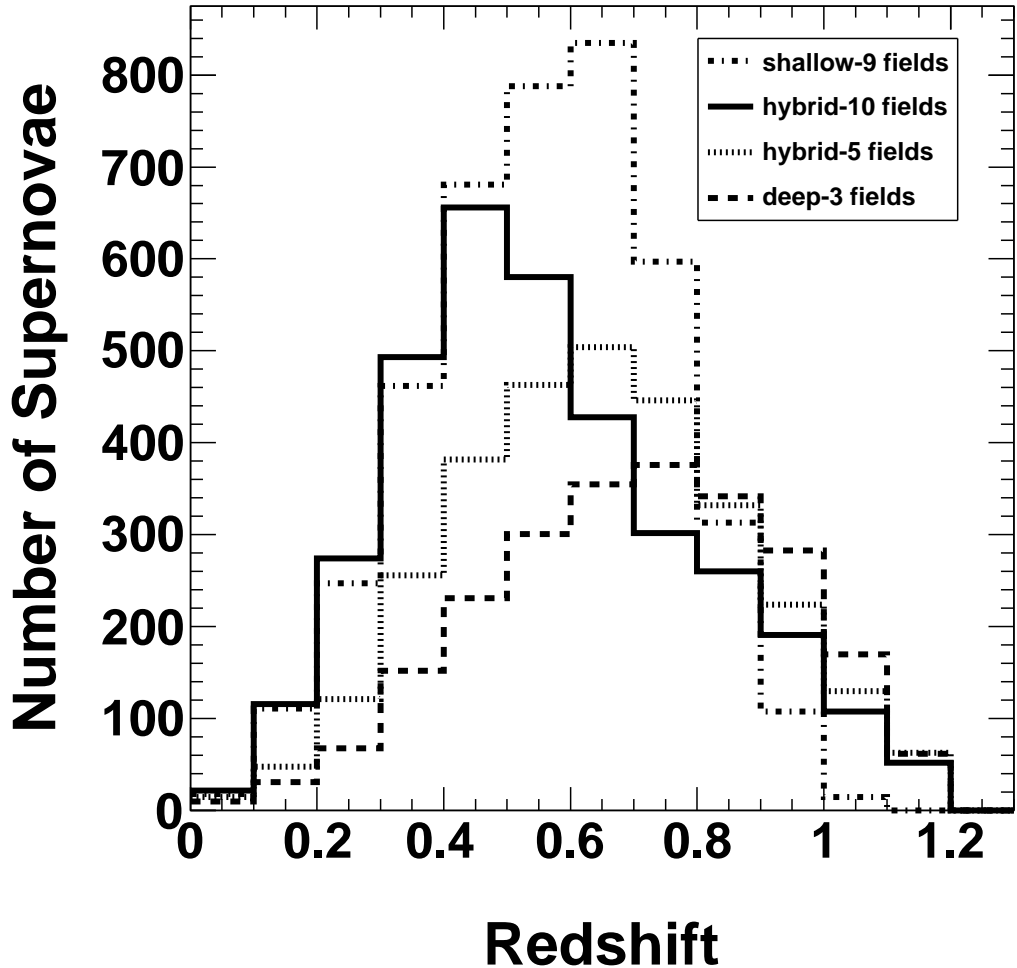

Fig. 10.—: Number of SNIa versus redshift for four of the DES strategies investigated. Total supernova statistics are 4175, 3482, 2984, 2381 for the shallow 9-field, hybrid 10-field, hybrid 5-field, and deep 3-field surveys respectively. The SN statistics shown include the application of all the selection cuts listed in Tab. 6. Note that subtle changes in the amount of exposure time allocated to each pass band can lead to large changes in the number of SNIa passing cuts. For example, a reasonable set of alternate exposure times considered for the 10-field hybrid results in ∼600 more SNIa passing cuts, mostly in the redshift range of 0.6-0.8. Such additional SNIa negate the apparent advantage of the 5-field hybrid survey in that redshift range as shown in this plot.

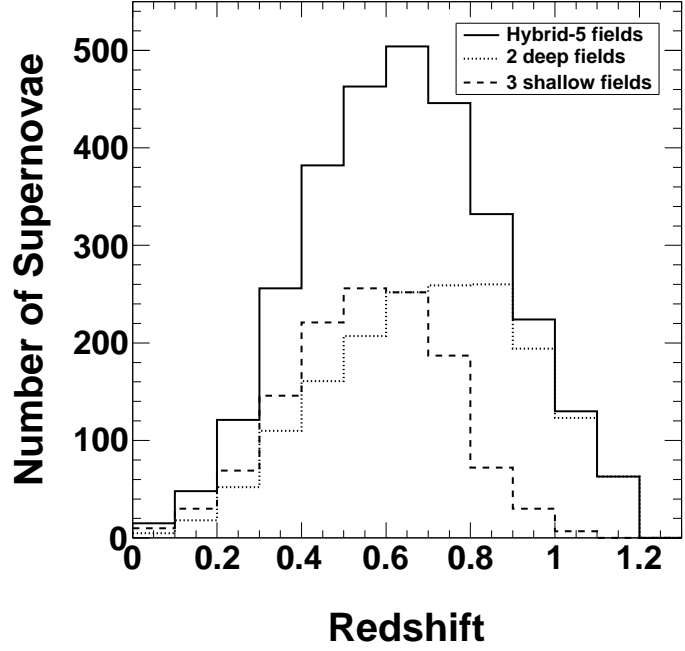

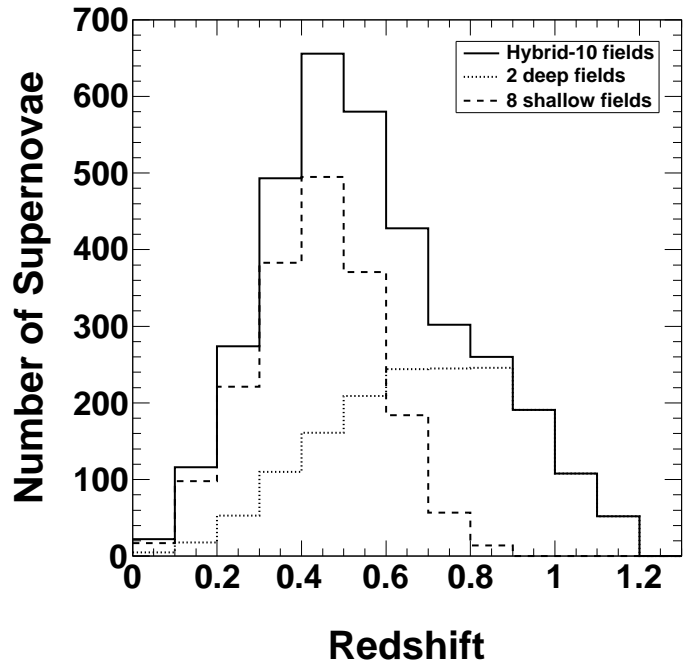

Fig. 11.—: *Top (bottom)*: the SNIa redshift distribution for the 5-field (10-field) hybrid survey including the deep and shallow components. Note that the 10-field cadence is slightly worse.

## 4. Redshift Determination

A precise estimate of SN redshifts is needed for placement of SNe on the Hubble diagram and for performing K-corrections on observed pass-bands to the SN rest frame. There are four possible methods of obtaining SN redshifts: 1) spectroscopic follow-up of individual SNe, 2) spectroscopic redshifts of the associated host galaxies, 3) photometric redshifts (photo-z's) of SNe, and 4) photo-z's of the host galaxies. In addition, the DES collaboration is considering the use of optical cross-correlation filters (Scolnic et al. 2009) for both redshift determinations and SN typing. The final analysis of the DES SNe will use the host spectroscopic redshifts as the central method for redshift determination, with important roles being played by the other methods. We next discuss the redshift determinations for the final analysis (with the complete sample of host galaxy spectra and redshifts), as well as the interim analysis before host spectroscopic redshifts have been measured.

### 4.1. Role of Each Method of Redshift Determination

In previous SNIa Hubble diagram analyses, cosmological constraints have been obtained using mostly spectroscopic confirmation of the SN, which not only afforded an extremely precise determination of the redshift, but also the additional advantage of accurate SN typing. For the DES, it is impractical to obtain spectra for every SN at high-z. The DES will use photometric typing for most of the SNe observed (see §6). This technique works very well, and will be further validated by obtaining a spectrum for a significant fraction of low-redshift SNe. In addition, a sample of $10-20\%$ of SNe at higher redshifts, with a spectrum taken with 6-10m class telescopes, will be used to study SN evolution, photo-z's, and sample purity. Note that SNe with host galaxies too dim to obtain a host spectrum are another sample that could trigger taking of a follow-up SN spectrum.

Obtaining spectroscopic redshifts of host galaxies, assuming correct host identification, yields precise SN redshifts. In addition, large numbers of the host galaxies can be measured simultaneously with a multi-object spectrograph (MOS). We will target every visible SN host, but we expect that the efficiency of obtaining a valid redshift will decrease significantly for galaxies dimmer than apparent i-band magnitude $m_i = 24$, as indicated by the followup of SNLS galaxies (Hardin et al., in preparation). For the purposes of our study, we have approximated the efficiency of obtaining a galaxy redshift as 100% for $m_i < 24$ and 0% for $m_i > 24$. For forecasting SNe analyses, as well as planning follow-up telescope resources, it is important to estimate the fraction of SN hosts with $m_i < 24$. Measurements of SNIa host magnitudes from SNLS (Hardin et al., in preparation) have large statistical uncertainties at the highest SNLS redshifts. Therefore, we have constructed a model described in Appendix A. This is a non-trivial task, however, given uncertainties in the SNIa rate dependencies on galaxy mass, luminosity, and type and of redshift evolution. Appendix A describes, in detail, our estimates of SNIa host galaxy brightnesses in redshift bins, and the sources of significant uncertainty at large redshift. A model estimate is shown in Tab. 7, where we present the fractions of SNIa host galaxies satisfying the apparent magnitude limit $m_i < 24$ for z-bin values from 0.1 to 1.2. Within the uncertainties, the data and model agree. In this study, we choose to use the model (since it lacks the statistical fluctuations of the data) to remove from our cosmology analysis SNe without a host spectrum by applying the stated fractions (for the 10-field hybrid strategy, this cuts out 429 SNe, mostly at high redshift). We will present the impact of this choice on cosmological constraints in §8.

| Redshift | SNLS Data | Model |
|---|---|---|
| 0.1-0.2 | 100% | 98% |
| 0.2-0.3 | 94.4% | 97% |
| 0.3-0.4 | 97.4% | 94% |
| 0.4-0.5 | 96.5% | 92% |
| 0.5-0.6 | 94.1% | 89% |
| 0.6-0.7 | 79.0% | 85% |
| 0.7-0.8 | 88.6% | 82% |
| 0.8-0.9 | 78.4% | 78% |
| 0.9-1.0 | 76.9% | 74% |
| 1.0-1.1 | 50.0% | 70% |
| 1.1-1.2 | N/A | 67% |

Table 7:: Measured (SNLS, Hardin et al., in preparation) and estimated percentages of SNIa host galaxies with $m_i < 24$ are tabulated. The model values are taken from the middle column of Tab. 19 from Appendix A. For both the data and model, the uncertainties grow from a few % at low redshift to $\pm 25\%$ for z>1.0.

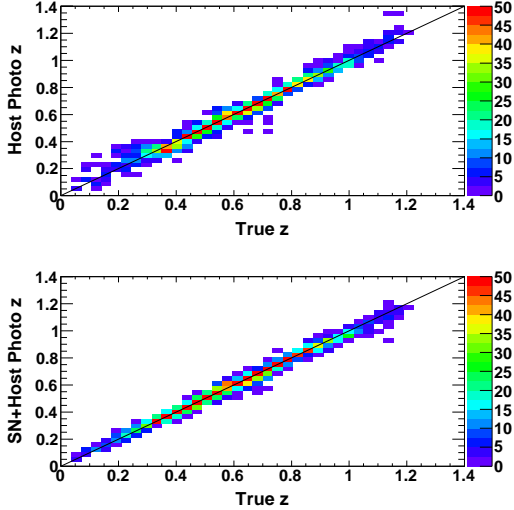

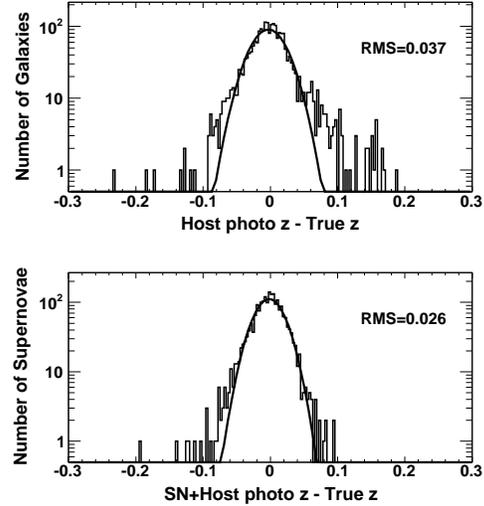

Fig. 12.—: Assuming the DES 5-field hybrid strategy, *top*: the estimated host galaxy photo-z is plotted versus the true redshift, with colors representing the number of SNe per bin; *bottom*: the SN photo-z (with the host galaxy photo-z used as a prior in the fit) is plotted versus the true redshift.

Fig. 13.—: Assuming the DES 5-field hybrid strategy, *top*: histogram of host galaxy photo-z minus the true redshift overlaid by a Gaussian ($\sigma = 0.027$) fit to the data, which was measured to have an RMS=0.037; *bottom*: histogram of SN photo-z (with the host galaxy photo-z used as a prior in the fit) minus the true redshift overlaid by a Gaussian ($\sigma = 0.022$) fit to the data, which was measured to have an RMS=0.026.

Photo-z's, both of the host galaxy and the SN, will play four roles in the DES: they will provide 1) interim SN redshifts before host galaxy spectra are available, 2) an opportunity to supplement the SN sample with redshifts if the host galaxy is dimmer than $m_i = 24$ or if the host spectrum cannot be obtained for other reasons, 3) a check on host galaxy identification when redshift comparisons are possible, and 4) help in classifying SNe during the search and in prioritizing them for spectroscopic follow-up. Two key elements of our photometric redshifts are: 1) a deep, ∼35 measurement co-add, per season, of the host galaxy, and 2) a combined SN+host photo-z fit using the host photo-z as a prior. In the next two sections, we show that the combination of these two elements give photo-z's the precision needed to play the roles in the DES SN analysis mentioned above.

### 4.2. Accuracy of photometric redshifts

The DES photo-z's will come from a combination of host galaxy photo-z and SN photo-z measurements. The host galaxy photo-z is expected to be relatively accurate since each SN field will be sampled more than one hundred times over the five-year survey. The limiting magnitudes of the SN host galaxy, 1-season co-add will be ∼26th mag, compared to ∼24th mag for the standard DES field. The limiting magnitude of a 5-season co-add will be ∼27th mag. DES expects to have at least 60K host galaxy spectroscopic redshifts for training photo-z's (H. Lin, private communication, 2011). In our simulations, the host galaxy photo-z is determined by a neural-net algorithm described in Oyaizu et al. (2008). Also from Oyaizu et al. (2008), the photo-z error is estimated by the Nearest Neighbor Error algorithm. Figs. 12 & 13 show scatter plots of photometric versus true redshifts for galaxies with a magnitude less than 24th and the histograms for the difference of host/SN pho-

### 4.3. Photometric redshifts for hosts without spectra

The second role for photo-z's is to supplement redshifts from host spectra at high-z, assuming the host spectra are only available for $m_i < 24$. We have prepared a simulated sample (detailed in Appendix A) of galaxy photo-z's that has been trained on a sample of $m_i < 24$ galaxies, but has then been applied to galaxies with $24 < m_i < 26$. Fig. 14 shows histograms of the host photo-z residuals from this sample, and the combined SNe+host photo-z. We will investigate the impact of using these photo-z's on a cosmology analysis in §8.

At this time, we are assuming that SNe with hosts dimmer than $m_i = 26$ will not be used in a cosmology analysis, although with a 5-season co-add it is likely that many of those hosts will be observed and may provide an interesting sample to study.

### 5. Supernova analysis with spectroscopic redshifts

In this section, we discuss SNIa analysis for the case where every SN has a spectroscopic host-galaxy redshift, and correct SNIa identification is assumed (see §6), with an emphasis on the extraction of distance estimates. In order to enhance the robustness of our results, we employ both the MLCS2k2 and SALT2 models to simulate and fit SN light curves. For MLCS2k2, we consider cases of fitting both with and without correct priors on host galaxy extinction.

#### 5.1. MLCS2k2 light curve fitting with full priors

The use of a prior on the MLCS2k2 extinction parameter $A_V$ improves the determination of the distance modulus when the measurement error on $A_V$ becomes wider than the width of the $A_V$ distribution. The improvement is noticeable in the simulated DES data at high redshifts where the SN colors are determined by measurements in only three bands: $r$, $i$, and $z$. However, the use of a prior is susceptible to the introduction of biases if implemented with incorrect information. While measurement errors, in principle, average to zero when the measurements of many SNe are combined, a tometric redshifts and true redshifts, respectively. The host galaxy photo-z's have a Gaussian sigma of ∼0.027 and a non-Gaussian tail. The SN photo-z is fit with SNANA, using the host galaxy photo-z as a prior (Kessler et al. 2010a), is seen to have a Gaussian sigma of ∼0.022 and much-reduced tails. When added to the spectroscopic redshifts provided by SN follow-up, these redshifts are precise enough to begin an interim analysis of DES SNe before host spectra are available.

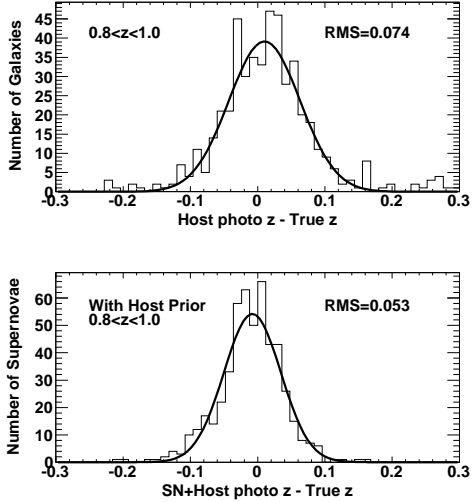

Fig. 14.—: Using simulated photo-z's trained on a sample with $m_i < 24$, but applied to a dimmer sample with $24 < m_i < 26$, the photo-z precision is presented for $0.8 < z < 1.0$ for the DES 5-field hybrid strategy. *Top*: Histogram of host galaxy photo-z minus the true redshift (RMS=0.074, $\sigma = 0.047$); note that there are 514 total entries with 19 underflows and 9 overflows. *Bottom*: histogram of SNe photo-z (with the host galaxy photo-z used as a prior in the fit) minus the true redshift (RMS=0.053, $\sigma = 0.042$). Similar histograms for $1.0 < z < 1.2$ demonstrate the following widths: Host galaxy only (RMS=0.09, $\sigma = 0.059$), SN with host prior (RMS=0.079, $\sigma = 0.045$); note that there are 514 total entries with 4 underflows and 0 overflows

bias in the prior will not average to zero. Inaccuracies in the prior, which is essentially the distribution of SNe in $A_V$, can arise from purely experimental errors. However, unknown astrophysics, including the evolution of the host galaxy population or the SN colors with redshift, pose serious challenges to the use of a prior in a high precision survey like the DES. While we do provide some estimate of potential systematic errors resulting from the use of a prior based on the SDSS analysis, our estimates must currently be considered preliminary.

For the analysis presented here, the prior has the following definition:

$$P_{\mathrm{prior}} = P(A_V) \times P(\Delta) \times \epsilon_{\mathrm{cuts}}(z, A_V, \Delta), \quad (2)$$

where $P(A_V)$ & $P(\Delta)$ are the underlying physical $A_V$ & $\Delta$ (luminosity parameter) distributions and $\epsilon_{\mathrm{cuts}}$ is the fraction of SNe that pass the selection cuts for a given redshift, $A_V$, & $\Delta$. For this work, following Kessler et al. (2009a), $P(A_V)$ is given by $dN/dA_V = exp(-A_V/\tau_{A_V})$ with $\tau_{A_V} = 0.334$, and $P(\Delta)$ is an asymmetric Gaussian with peak position, $\Delta_0$, and positive and negative side widths, $\sigma_+$ and $\sigma_-$, respectively, given by $\Delta_0 = -0.24$, $\sigma_+ = +0.48$, $\sigma_- = +0.23$. In addition, we set $dN/dA_V=0$ for $A_V<0$.

For a given survey, e.g., the 5-field DES hybrid scenario, $\epsilon_{\mathrm{cuts}}$ is calculated using `SNANA` by cyclically simulating SN light curves and checking which light curves pass the defined selection cuts until the desired efficiency accuracy is reached. Fig. 15 shows the selection efficiencies for various classes of SNIa. Both the deep and shallow observation fields within the 5-field hybrid survey exhibit statistical completeness for nearby and/or bright SNe. However, Fig. 15 shows the vastly higher efficiency of the deep relative to the shallow fields for distant and faint and/or heavily extincted SNe. Figure 16a shows our application of efficiencies to the hybrid survey simulation in order to avoid the bias in the fitted distance modulus that would arise from `MLCS2k2` light curve fitting with an incorrect prior, e.g., one with the assumption of a flat prior on efficiency.

As discussed above, the introduction of priors can easily lead to biases if the effects of the survey selection efficiency are poorly understood. Of particular concern is the bias manifested as a difference between observed (i.e. "fitted") and true (i.e., "simulated") distance modulus ("$\mu_{\text{fit}}$" and "$\mu_{\text{sim}}$" hereafter) that can arise. Figure 16f shows such a departure of $\mu_{\text{fit}} - \mu_{\text{sim}}$ from zero beyond a redshift of $\sim 0.7$. The bias illustrates the size of the $\mu$-correction that the DES SN data would need if the efficiency prior were incorrectly assumed to be flat. Note, one does not expect the selection bias to have a significant effect at low redshift because there the SN sample is essentially complete. The fact that $A_V$ is driven toward zero, while the trend in $\Delta$ is negative, as redshift increases beyond $\sim 0.5$ (see Fig. 17), implies that only less extincted and/or brighter SNe pass the selection cuts, and strongly supports our identification of the bias in $\mu$ as a selection bias. In addition, this selection effect explains the small drop in RMS beyond a redshift of z=1.0 exhibited in Fig. 16a. Figure 16d also shows, when compared to Fig. 16f, one of the key motivations for the hybrid survey. A systematic check is enabled by the ability to compare the less biased distance moduli from the deep part of the dataset at higher redshifts with the more biased shallow part. If that crosscheck is validated, then confidence is increased in the highest region of the deep component of the survey (i.e., redshifts greater than 1.0) where the deep component suffers a similar bias to that experienced by the shallow component at intermediate redshifts. Fig. 16e, showing the case of the deep-only strategy, is included for completeness.

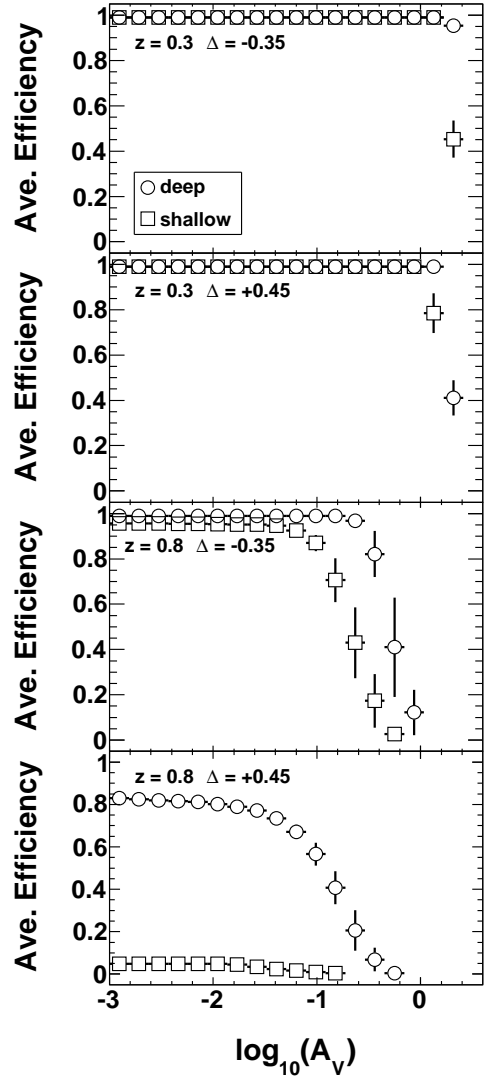

Fig. 15.—: Plotted from top to bottom is the efficiency due to the selection cuts discussed in §3.1 as a function of the extinction parameter, $A_V$, for the DES deep and shallow fields assuming the 5-field hybrid strategy. The efficiencies were calculated to an accuracy of 1% for a given redshift and value of $\Delta$, $A_V$, and $R_V$. The vertical error bars show the range in efficiency for an extreme variation in $R_V$ from 0.5 to 4.00 in a given $A_V$ bin. For the purposes of this plot, the pre- and post-epoch cuts were disabled. This was done in order to show the efficiencies without edge effects which reduce the peak efficiencies by approximately 10-15% for the cases in the top three panels.

## 5.2. `MLCS2k2` light curve fitting with flat priors & `SALT2` fitting

In this section, we discuss `MLCS2k2` flat-prior and `SALT2` model fitting. Such fits avoid the issue of selection efficiency bias discussed above. The trade-off is an increase in the RMS spread in the distance modulus, as is clearly evident in the comparison of Fig. 16a with Fig. 16b. In addition, Fig. 16b shows a high-redshift $\mu$ bias evident in `MLCS2k2` fits with flat priors. This is due to the fact that such fits allow negative values of $A_V$, for which the fitter compensates by pulling the distance modulus to higher values.

The `SALT2` light curve fitter in `SNANA` is accompanied by a separate program called `SALT2mu` (Marriner et al. 2011) that fits the `SALT2` parameters $\alpha$ and $\beta$ that are used to determine the standard SNIa magnitudes. The parameters that correlate distance modulus with $x_1$ (a stretch-like parameter) and $c$ (the color) are $\alpha$ and $\beta$ respectively. We have chosen to fit for the $\alpha$ and $\beta$ parameters independent of the cosmology using `SALT2mu`, which allows us to apply the same cosmological fitting procedure to the outputs of the `MLCS2k2` and `SALT2` light curve fits.

The resulting distance modulus residuals are shown in Fig. 16c. The trend in the RMS spread of the distance modulus is rather similar to that obtained with `MLCS2k2` with the use of a flat prior. While it would be possible to apply a prior on the color in the `SALT2` fit, we have followed normal practice in not doing so here. For the remainder of this paper, we will use `MLCS2k2` fits with correct priors (corresponding to Fig. 16a) in our analysis, with the exception that we use `SNooPy` in §7.2 and include `SALT2` in the discussion of the DES SN cosmology fits in §8.

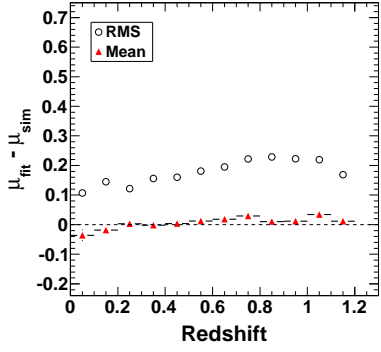
(a) `MLCS2k2` fit for the 5-field hybrid strategy with correct priors (see Eqn. 2).

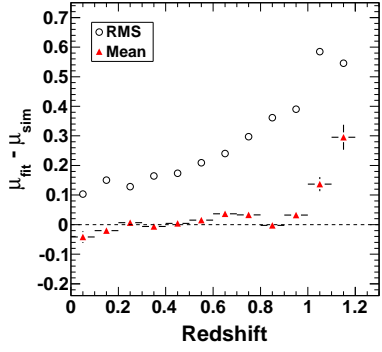
(b) `MLCS2k2` fit for the 5-field hybrid strategy with flat priors.

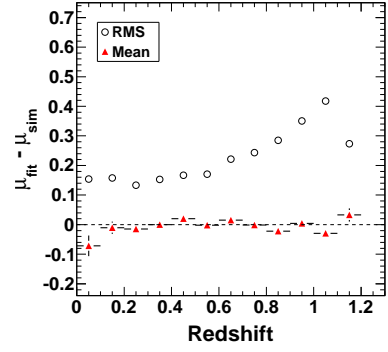
(c) `SALT2` fit for the 5-field hybrid strategy.

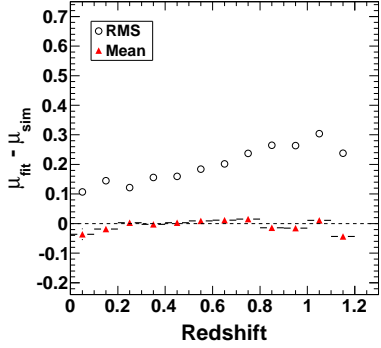
(d) `MLCS2k2` fit for the 5-field hybrid strategy using a prior based on the underlying $A_V$ distribution but not the simulated efficiency (see Fig. 15 for example efficiencies).

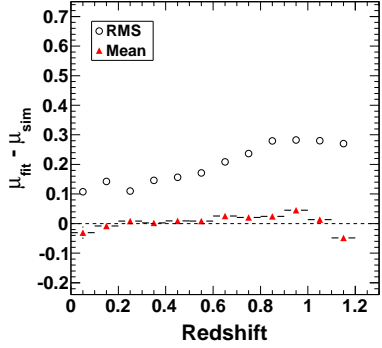
(e) `MLCS2k2` fit for deep strategy using a prior based on the underlying $A_V$ distribution but not the simulated efficiency.

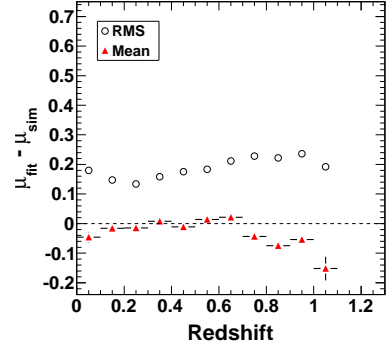
(f) `MLCS2k2` fit for shallow strategy using a prior based on the underlying $A_V$ distribution but not the simulated efficiency.

Fig. 16.—: Plotted is the fitted distance modulus residual ($\mu_{\rm fit}$ - $\mu_{\rm sim}$) for different SN light curve fitting scenarios. Dashed lines are drawn at zero for clarity.

## 6. Type Ia supernova sample purity

Since the DES SNIa sample will not have full spectroscopic SN follow-up, cases where core-collapse SNe (SNcc) are misidentified as SNIa will be a concern for a cosmology analysis based on the full sample. In order to address this issue, we have undertaken an analysis of the DES SNIa sample purity using `SNANA` simulations. In this study, we perform a mock-analysis using redshifts determined from the visible host galaxies. We have limited measurements of SNcc types, rates, and brightness, but our knowledge of SNcc is lacking in several areas, as discussed in detail below. There are substantial uncertainties in the absolute rate of SNcc, mean absolute magnitudes and their variance, relative fractions of the different types of SNcc, and variation in the light curve shapes that are not adequately represented in the simulation. This section will address these uncertainties and provide estimates of their effect on SNIa sample purity. In general, where there are choices to be made, we make the choice that will increase the amount of misidentification in order to see the worst-case effect on a cosmology analysis, as discussed in §8.

### 6.1. Core collapse input rate

In order to simulate SNcc, we use the input SN rate parametrization of Dilday et al. (2008), which found the SNIa rate from SDSS to be of the form $\alpha(1+z)^\beta$ with $\alpha_{Ia} = 2.6\times10^{-5}$ with $\alpha_{Ia} = 2.6 \times 10^{-5}$ SNe $h_{70}^3$ Mpc$^{-3}$ yr$^{-1}$, and $\beta_{Ia} = 1.5$. For SNcc, we take $\beta_{cc} = 3.6$ to match the star formation rate. Various studies, the most recent being SNLS (Bazin et al. 2009), have shown this assumption to be valid, albeit with low statistics and limited redshift range. This leaves the determination of $\alpha_{cc}$. Taking the ratio of SNcc/Ia to be the SNLS value of 4.5 for redshifts of $< 0.4$ (Bazin et al. 2009), we calculate the value $\alpha_{cc}$ must have in order to obtain the ratio of 4.5: $\alpha_{cc} = 6.8 \times 10^{-5}$ SNe $h_{70}^3$ Mpc$^{-3}$ yr$^{-1}$. Note that with this value of $\alpha_{cc}$, the SNcc/Ia ratio increases to $\sim 10$ out to a redshift of 1.2. A caveat in this estimate is that one of the largest uncertainties is the actual population near the detection threshold. Direct measurements of the SNcc rate beyond a redshift of z=0.4 would be very helpful in the determination of SNIa sample purity.

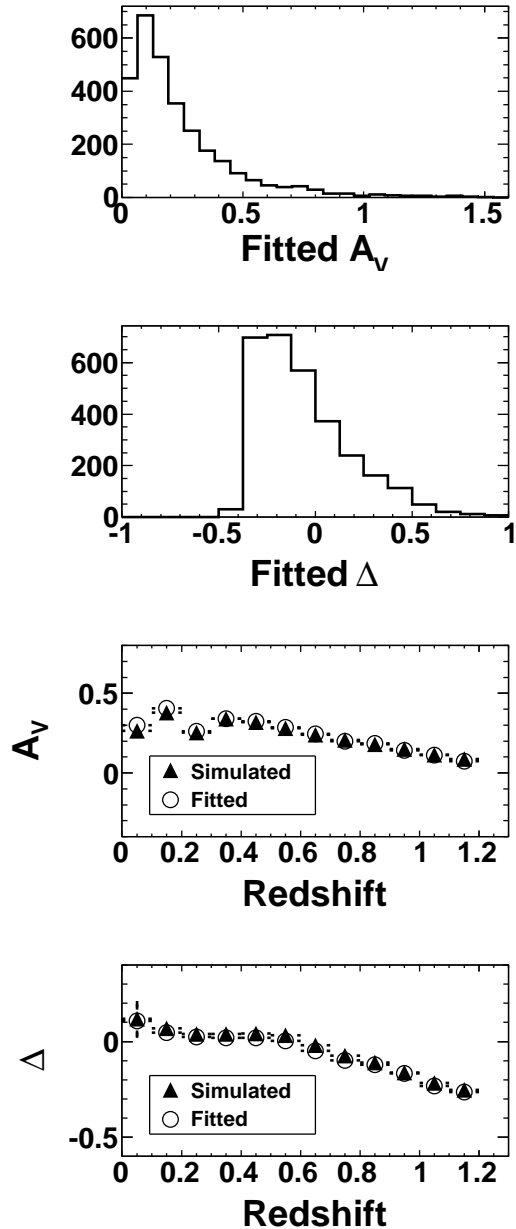

Fig. 17.—: Plotted from top to bottom is the fitted $A_V$ histogram, the fitted $\Delta$ histogram, the redshift dependence of simulated & fitted $A_V$, and redshift dependence of simulated & fitted $\Delta$, both averaged within a redshift bin, assuming the 5-field hybrid strategy. Note that the lowest redshift bin has low SN statistics (see Fig. 10).

## 6.2. Relative fractions of core collapse types

In this section, we discuss the relative fraction of the SNcc subtypes. The most important fraction is that of Type Ib/c, since they most commonly pass the combination of cuts on SNRMAX and MLCS2k2 fit-probability that the SN is a SNIa ($f_p = P_{\chi^2}$, the probability from fit $\chi^2$ and the number of the degrees of freedom). The literature contains several estimates of the ratio of Type Ib/c to Type Ib/c plus II SNe (see Tab. 8 for examples). The most complete references, in terms of fractions being given for each type of SNcc, are Li et al. (2011a) and Smartt et al. (2009), and the Type Ib/c fractions are in good agreement. We have used the Smartt et al. (2009) values (see Tab. 9) as the default set of fractions in this analysis, as they give a more conservative amount of SNcc misidentification relative to Li et al. (2011a).

| Reference | Ib/c fraction |
|---|---|
| Li et al. (2011a) | $24.6 \pm 4.6\%$ |
| Li et al. (2007) | $26.5 \pm 5.4\%$ |
| van den Bergh et al. (2005) | $24.7 \pm 2.6\%$ |
| Smartt et al. (2009) | $29.3 \pm 4.7\%$ |
| Prieto et al. (2008) | $24.7 \pm 4.9\%$ |
| Leaman et al. (2011) | $33.3 \pm 4.3\%$ |

Table 8:: References for the relative fraction of Type Ib/c SNe (number of Type Ib/c divided by the total number of SNcc of all types).

| SN Type | Relative SNcc Fractions |
|---|---|
| IIP | $0.587 \pm 0.05$ |
| Ib/c | $0.293 \pm 0.05$ |
| IIL + IIb | $0.082 \pm 0.03$ |
| IIn | $0.038 \pm 0.02$ |

Table 9:: The relative fraction of collapse SNe subtypes (number of a given subtype divided by total number of SNcc of all types) used in this analysis, as taken from Smartt et al. (2009).

## 6.3. Core collapse brightness

The absolute brightness of SNcc is a critical parameter in the number of SNcc misidentified as SNIa, since most are too dim to pass typical SNRMAX cuts (e.g., those shown in Tab. 6). Two references for absolute SNcc brightnesses, Richardson et al. (2002) and Li et al. (2011a), are compared in Tab. 10 and Tab. 11. The numbers in Tab. 10 have been corrected for the significant Malmquist bias evident in that data. The correction assumed a threshold of 16 magnitudes in apparent brightness, and took into account the larger volume sampled by intrinsically brighter SNe than for fainter SNe. The volume-limited analysis in Li et al. (2011a) is already corrected for Malmquist bias, but Conley et al. (2011) used Richardson et al. (2002) in their analysis, noting that Li et al. (2011a) perhaps missed a bright Type Ib/c component by avoiding low-luminosity galaxies. To take the conservative approach, we used the single-Gaussian-approximation brightnesses from Richardson et al. (2002) as our default.

| Richardson et al. (2002) | | |
|---|---|---|
| SN Type | $M_B$ | $\sigma_{M_B}$ |
| IIP | $-14.40 \pm 0.42$ | 0.81 |
| Ib/c | $-16.72 \pm 0.23$ | 0.62 |
| IIL | $-17.19 \pm 0.15$ | 0.47 |
| IIn | $-17.78 \pm 0.41$ | 0.74 |

Table 10:: The absolute B-band magnitudes and widths for the single Gaussian fits from Richardson et al. (2002), corrected for Malmquist bias.

| Li et al. (2011a) | | |
|---|---|---|
| SN Type | $M_R$ | $\sigma_{M_R}$ |
| IIP | $-15.66 \pm 0.16$ | 1.23 |
| Ib/c | $-16.09 \pm 0.23$ | 1.24 |
| IIL | $-17.44 \pm 0.22$ | 0.64 |
| IIn | $-16.86 \pm 0.59$ | 1.61 |

Table 11:: The absolute R-band magnitudes and widths from Li et al. (2011a).

## 6.4. Core collapse templates

SNcc are observed to be a much more heterogeneous class than SNIa and, in contrast to SNIa, there is no parametrization available that describes the diversity of SNcc light curves. Therefore, we take a template approach to modeling SNcc. Three sets of templates are compared, with each being a spectral sequences as a function of time. The first set are 40 templates from the Supernova Photometric Classification Challenge (Kessler et al. 2010b), the second set are the composite spectral templates constructed by Nugent[17], and the third set are Type Ib/c and IIP templates from Sako et al. (2011), augmented by the Nugent templates for Types IIL and IIn. All templates were converted to SDSS filter magnitudes, and SNANA performs the K-corrections into the DES filters. Note that there is no template for Type IIb, which in Li et al. (2011a) is more numerous than Types IIL or IIn. The Type IIL template is expected to be the closest to Type IIb SNe, and, therefore, we used it for the Type IIb sub-sample.

In the SNANA simulation, the templates are corrected to the absolute brightnesses discussed in the previous section. In addition, the Nugent templates are composite spectra and do not include absolute brightness fluctuations, therefore they are also smeared by the Gaussian-fitted widths tabulated in the previous section in order to better reflect the observations. The Kessler et al. (2010b) and Sako et al. (2011) templates already have sufficient variation in brightnesses and require no additional smearing. The templates from the Supernova Photometric Classification Challenge are the most complete set and are used as the default in the rest of this paper. In particular, note that these templates contain a "1+$z_{temp}$" bug in that each SNcc template is too dim by a factor of 1+$z_{temp}$ (see Tab. 4 and §2.6 of Kessler et al. 2010b), where $z_{temp}$ is the redshift of the template. In the next section, we include a discussion of the effect of this bug on our simulations.

### 6.5. Sample purity results

Using the inputs discussed above, and spectroscopic host redshifts, we simulated the DES SN sample including SN Types Ia, Ib/c, IIL, IIn, and IIP subject to the selection criteria listed in Tab. 6. The SNe in this combined sample are fit to the SNIa MLCS2k2 model, giving a fit probability $f_p$ variable cut that can be customized for each analysis, and for the amount of SNcc observed in a sub-sample with spectroscopic follow-up. Figure 18 shows the distribution of $f_p$ for the SNIa and SNcc samples, after all other selection cuts have been applied. The number of SNe of each type with no $f_p$ cut, and with $f_p > 0.1$, is shown in Tab. 12. For those results, the effect of the 1+$z_{temp}$ SNcc template bug discussed at the end of the previous is an increase in the SNIa purity by ∼2%, which has no impact on our conclusions. Figures 19 and 20 show redshift distributions of these samples subject to a fit probability cut $f_p > 0.1$. Table 13 shows comparisons in the total SNcc number with variations in the simulation inputs discussed above, with a range of ×3 in total sample SNcc. Note that the sample purity is better than that obtained with the same analysis performed in Kessler et al. (2010b); this is due to the correction for Malmquist bias applied to the core collapse simulation sample, which reduces their expected absolute brightness and therefore the number passing SNRMAX cuts.

| Sample | $f_p > 0.0$ | $f_p > 0.1$ | Tot. simulated |
|---|---|---|---|
| Ib/c | 571 | 57 | 53514 |
| IIP | 110 | 2 | 107210 |
| IIn | 225 | 2 | 6940 |
| IIL | 62 | 2 | 14976 |
| Tot. SNcc | 968 | 63 | 182640 |
| Ia | 3482 | 3350 | 18695 |
| Ia+SNcc | 4450 | 3413 | 201335 |
| Ia Purity | 78% | 98.1% | n/a |

Table 12:: Number of simulated SNe passing cuts and sample purity using the DES 10-field hybrid strategy for SNIa fit probability, $f_p$ cuts of 0.0 and 0.1. Note that employing $f_p > 0.2$ reduces the number of SNIa and SNcc passing cuts by 5% and 46%, respectively. However, given that the impact of SNcc on the DES cosmological constraints is already negligible assuming $f_p > 0.1$ (see §8), opting for $f_p > 0.2$ is unwarranted due to the loss of SNIa. Note that these results were obtained with SNANA v8_37, which includes a known bug due to each SNcc template being too dim by a factor of 1+$z_{temp}$ (see §2.6 of Kessler et al. 2010b), where $z_{temp}$ is the redshift of the template. We have verified that employing fixed versions, e.g., v9_89, results in a small (∼2%) purity variation that does not have an effect on our conclusions.

---

[17] http://supernova.lbl.gov/%7Enugent/nugent_templates.html; see also Nugent et al. (2002).

| Simulation Input | Total SNcc |
|---|---|
| Defaults | 63 |
| Nugent templates | 27 |
| Sako et al. templates | 44 |
| Li et al. abs. magnitudes | 8 |

Table 13:: Total SNcc counts with variations in the simulation inputs assuming the 10-field hybrid strategy, with $f_p > 0.1$. The line labeled "Defaults" is the same as the Total SNcc in Tab. 12.

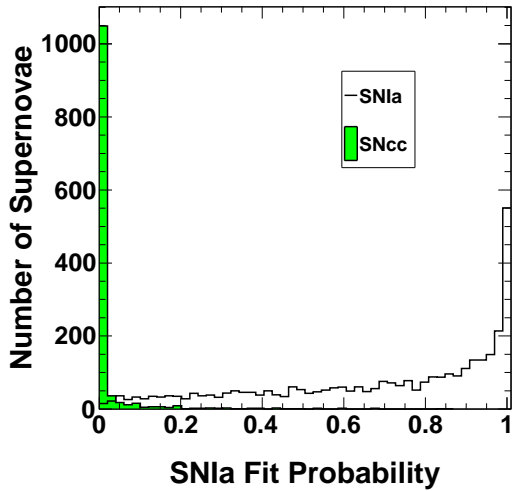

Fig. 18.—: Plotted are the SNIa fit probabilities for the SNIa and SNcc samples, after all other selection cuts are applied.

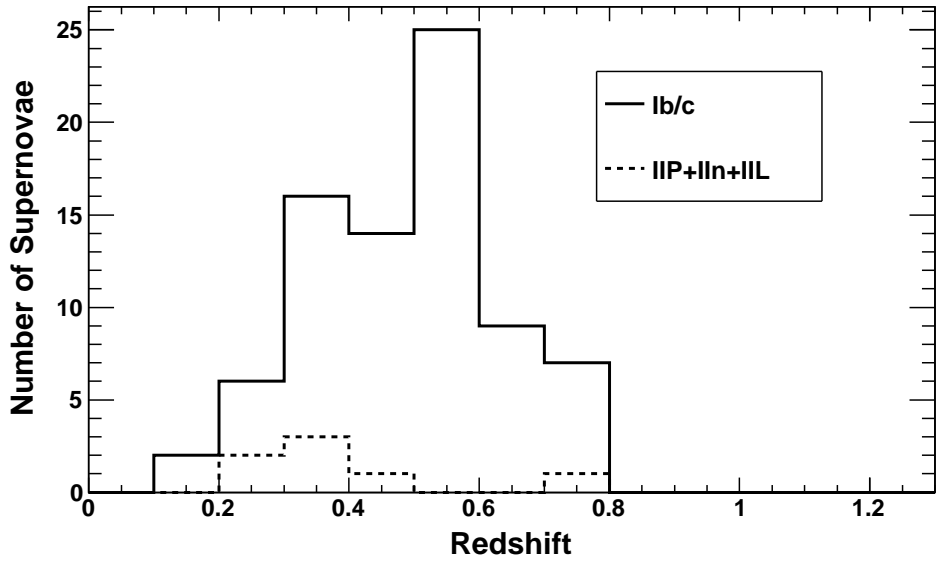

Fig. 19.—: Plotted are the histograms showing the projected DES redshift distributions for the Type Ib/c SNe and the summed distribution of other core collapse SNe, assuming the 10-field hybrid survey, the selections criteria in Tab. 6, and $f_p > 0.1$.

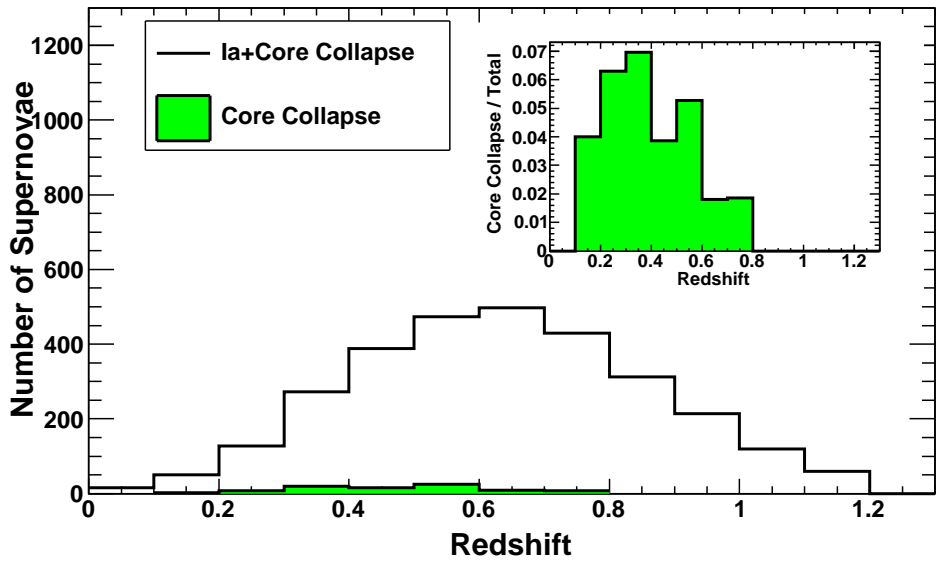

Fig. 20.—: Plotted are the redshift distributions for the projected DES SNIa and non-Ia SN samples assuming the 10-field hybrid strategy, the selections criteria in Tab. 6, and $f_p > 0.1$.

## 7. Supernova colors, dust extinction, and infrared data

The study of SN colors is a rich subject that is of crucial importance to SN cosmology. The issue of confusion between intrinsic color variations and dust extinction, which complicates the measurement of the former, is beyond the scope of this paper. Instead, we demonstrate the DES sensitivity to variations of the traditional, redshift-independent dust parameters $A_V$ and $R_V$ (Cardelli et al. 1989). Color measurements in the DES will be improved by the enhanced red sensitivities of the CCDs, as discussed in §1.

for a grid of values of $R_V$ and $\tau_{A_V}$, where $\tau_{A_V}$ is the parameter that controls the width of the simulated $A_V$ distribution, as described in §5.1. As $\tau_{A_V}$ increases, the $A_V$ distribution extends to larger extinctions and, thus, produces SNe with redder colors. Our reference color sample is a simulation with the values of $R_V = 2.18$ and $\tau_{A_V} = 0.334$, which are the best fit values from Kessler et al. (2009a). The results presented here are for the redshift range $0.4 < z < 0.7$. This range has the highest SN statistics for the DES. For the redshift range $z < 0.4$, the SN statistics are much less, but SNRMAX is substantially better, so that the precision of the color measurements are comparable to those presented here.

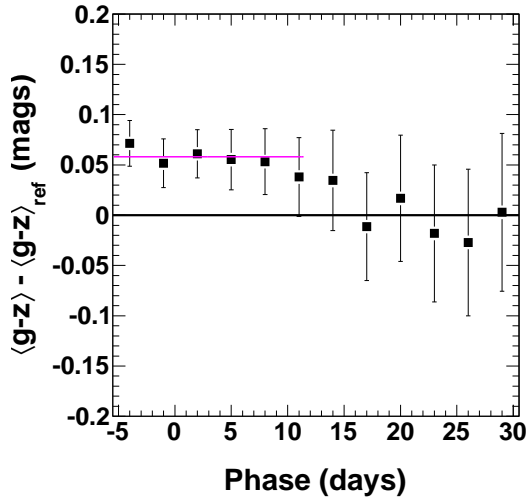

Fig. 21.—: Average DES $g - z$ color difference versus phase assuming the 5-field hybrid strategy for a simulation with $R_V = 2.69$ and $\tau_{A_V} = 0.25$, as compared to the reference simulation with $R_V = 2.18$ and $\tau_{A_V} = 0.334$. Error bars are the error on the mean color difference. The solid, horizontal line above zero shows the fitted average $g - z$ color difference for phase $< +11$ days. Note that since the quantity plotted a difference between colors, the errors, which are the quadrature sum of the errors on the mean of each color, are correspondingly large.

### 7.1. Sensitivity to $A_V$ and $R_V$

We perform an analysis of the color variations in the SN colors $g - i$, $g - z$, $r - i$, and $r - z$

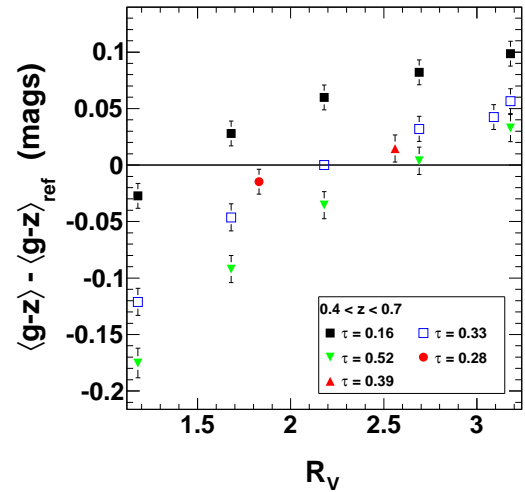

Fig. 22.—: Average DES $g - z$ color difference assuming the 5-field hybrid strategy for phase $< +11$ days compared to the reference simulation with $R_V = 2.18$ and $\tau_{A_V} = 0.334$ as a function of $R_V$ and for a range of $\tau_{A_V}$. Error bars are the error on the mean color difference. Note the isolated points for $\tau_{A_V} = 0.28$ and $0.39$. We use these points to set the values of the $1\sigma$ errors in $R_V$ and $\tau_{A_V}$ to be 0.38 and 0.06, respectively.

We constructed a suite of simulations with a grid of $R_V$ and $\tau_{A_V}$ values in order to assess the effects of changes in $R_V$ and $\tau_{A_V}$ on SNIa colors. An example of the effects on the $g - z$ color is shown in Fig. 21. The differences in color between simulations with $R_V = 2.69$ and $\tau_{A_V} = 0.25$ and our reference sample parameters is shown as a function of phase. A signal-to-noise cut of 0.5 is applied at every phase. Fig. 21 has two noteworthy features: the fitted average color level for phase $< +11$ days and the significant drop in color for later phases. The average color level of a given SN color for phase $< +11$ days has, in general, a complicated dependence on $R_V$, $\tau_{A_V}$, and the redshift range of the data sample. In special cases, for simulations with fixed $R_V$, $A_V$, and certain values of fixed redshift, this dependence can be predicted from the CCM dust model (Cardelli et al. 1989) and the parametrization from Jha et al. (2007). We have verified that our simulations agree well with the predictions in these cases.

The sensitivity to parameters $R_V$ and $\tau_{A_V}$ of the small-phase average $g - z$ difference is shown in Fig. 22. The error bars show the statistical uncertainty for each parameter choice. Overall, the trend is to increase the value of the $g - z$ difference by approximately 0.3 magnitudes as $R_V$ increases from 1.1 to 3.1, which is a plausible range for $R_V$, and $\tau_{A_V}$ increases from 0.16 to 0.52. From this figure, simulated values of $R_V$ within $\sim 0.38$ of the reference value, and of $\tau_{A_V}$ within $\sim 0.06$ of the reference value, can be distinguished at the $1\sigma$ level. Similar plots for other SN colors and other redshift ranges show slightly different dependencies on the parameters, and hence can be used to lower further the above uncertainties in $R_V$ and $\tau_{A_V}$. Fig. 22 also shows several degenerate combinations of $R_V$ and $\tau_{A_V}$ that lead to the same level of $g - z$ difference. This occurs because SN colors are largely dependent only on the ratio of $A_V$ to $R_V$, and so a given color difference can only determine the ratio of $A_V$ to $R_V$ to some uncertainty. This degeneracy is reduced by considering the behaviors of other color differences and their redshift dependence. In addition, the second feature of Fig. 21, namely the drop-off in the $g - z$ difference at late phases[18], can also be used to resolve this degeneracy. In this analysis, we assume that the degeneracy can be broken by an SDSS-like analysis (Kessler et al. 2009a), which took all such effects into consideration. Therefore, in our analysis in §8, we take the uncertainty in $R_V$ and $\tau_{A_V}$ to be 0.38 and 0.06, respectively.

---

[18]This drop-off is due to an effect of the SNRMAX cut: for redshifts greater than $z \approx 0.4$, where the DES is no longer fully efficient, the SNRMAX cut is more likely to remove the fainter, redder SNe at late phases.

## 7.2. VIDEO survey and additional infrared overlap

### 7.2.1. The DES+VIDEO overlap

The infrared VISTA Deep Extragalactic Observations (VIDEO) Survey (see, e.g., Jarvis 2009), using the Visible and Infrared Survey Telescope for Astronomy at the Paranal Observatory in northern Chile, began science observations in late 2009. This 5-year survey has an area of 12 deg$^2$ covering 4.5 deg$^2$ in XMM-LSS, 4.5 deg$^2$ in *Chandra* Deep Field South, and 3 deg$^2$ in ELAIS S1, with deep observations in the $Z$, $Y$, $J$, $H$, $K_s$ filter set. The survey is designed to trace galaxy evolution out to a redshift of 4, and also provides for a large-volume SN search projected to find 250 SNcc and 100 SNIa with a median redshift of 0.2.

The VIDEO Survey SN fields overlap those for the DES (see Tab. 2). The extension of optical SNIa light curves to include infrared data points enables an enhanced determination of SN colors and dust extinction due to the larger lever arm provided by the increased wavelength range. As emphasized by Freedman et al. (2009), which presented the first $i$-band Hubble diagram obtained by the Carnegie SN Project, infrared SN observations offer advantages in reducing several systematic effects, the most notable of which is reddening due to dust. In particular, near-infrared observations can be used to obtain a SN data set that is insensitive to variations in SN color, and therefore facilitate the best rate assessments for different SN types and their dependence on host galaxy properties. In order to simulate expected results from a combined DES+VIDEO dataset, we incorporated the optical+infrared `SNooPy` SN light-curve model (Burns et al. 2011) into `SNANA`. Such a dataset, even with modest SN statistics, enables the pursuit of reduced-extinction systematics studies.

### 7.2.2. The DES+VIDEO supernova sample

Based on VIDEO Survey SN data from the first season, we constructed a `SNANA` simulation library (see §2.2) with the following characteristics: typical $Y$- and $J$-band PSF is of order 1 arcsec, sky noise on the order of 200-400 photoelectrons, and zeropoints ranging from 31.5 to 32.0 magnitudes. Two seasons of VIDEO/DES overlap are expected, and the simulation library assumes that the observing conditions will be similar during both sea-sons. In addition, there are 10 observations in $Y$-band and 13 in $J$-band, each with 32 minutes of exposure time. Based on this simulation library, we estimate that the DES+VIDEO combined SNIa sample from years 2013 and 2014 could consist of approximately 108 SNe with z<0.5 in the common *Chandra* Deep Field South, XMM-LSS, and ELAIS S1 fields (see Tab. 2). Figures 23 and 24 show VIDEO SNRMAX for the 108 overlapping SNe, and an example simulated combined light curve, respectively. As shown in Fig. 23, some of the SNe have SNRMAX that are relatively low (e.g., < 5), and may not be useful for all SNe analyses. As a follow-on to this analysis, a study is planned to utilize `SNANA` simulated DES+VIDEO SNIa light curves to evaluate the benefit to SN color determinations to be gained by adding VIDEO infrared SNIa data to the DES SN analysis.

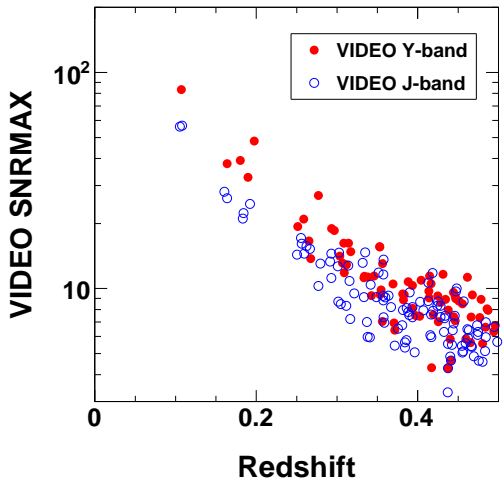

Fig. 23.—: SNRMAX as a function of redshift for the VIDEO $Y$- and $J$-bands.

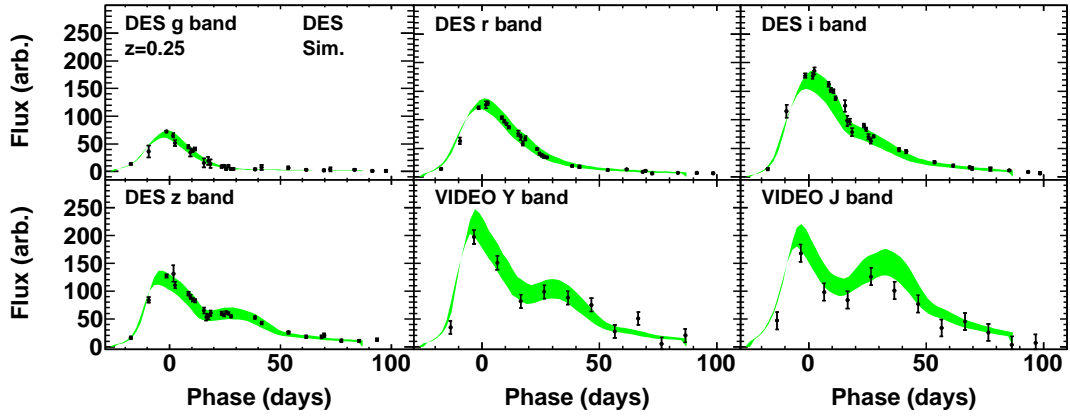

Fig. 24.—: Example simulated Type Ia SN light curve forecast displaying combined DES and VIDEO Survey data assuming the DES 5-field hybrid strategy. The points are a `MLCS2k2` fit and the band is the fit error. The wavelength ranges in nanometers of the pass-bands indicated are: 400–550 ($g$-band), 560–710 ($r$-band), 700–850 ($i$-band), 850–1000 ($z$-band), 970–1020 ($Y$-band), 1040–1440 ($J$-band). Note that initial investigations show that the $H$ and $K_s$ SNRMAX is insufficient for SN science, and so only a $grizYJ$ light curve is shown.

## 8. Dark energy constraints from different survey Strategies

In this section, we present a forecast of the constraints on cosmological parameters from the DES SN search using our simulations of the different survey strategies summarized in Tab. 3 (with the exception of the ultra-deep strategy, which is not considered here). In order to be included in the analysis in this section, each SN is required to pass the selection criteria listed in Tab. 6. In order to ensure an accurate spectroscopic host galaxy redshift determination, SNe with faint hosts ($m_i < 24$) are discarded. A multi-color light curve fit for each SN in the sample is made using the MLCS2k2 model with the prior listed in Eqn. 2, as described in §5. In this section, we also include results using the SALT2 model for comparison. A SNIa fit probability cut of 0.1 ($f_p > 0.1$) is made to reject SNcc, as described in §6. We make the forecasts in the context of the CPL parametrization (Chevallier & Polarski 2001; Linder 2003) of the dark energy equation of state, $w(a) = w_0 + (1-a)w_a$. The cosmological parameters relevant for SN observations that we included are $\Omega_{DE}, w_0, w_a, \Omega_k$, which are the dark energy density, the dark energy equation of state parameters, and the spatial curvature parameter. A typical binned Hubble diagram for DES SNe, from the 10-field hybrid survey, is shown in Fig. 25. The line represents the flat $\Lambda CDM$ cosmology calculation used in the simulation (as described in §2). The RMS scatter for the binned DES SNe, as well as RMS/$\sqrt{N}$, is shown in Tab. 14. The small drop in RMS for redshifts beyond z=1.0 was discussed in §5. Table. 14 completes the picture by showing that RMS/$\sqrt{(N)}$ continues to increase at the highest redshift, as expected. Figure 26 is another version of the Hubble diagram, this time with individual SNIa and SNcc for the hybrid 10-field survey. The Hubble diagrams for the hybrid 5-field survey are very similar to the 10-field figures. These figures will be discussed further in §8.2.

The DES SN sample will provide the most precise cosmological constraints when combined with low-redshift SN samples. We include in our forecasts a simulation of the 3-year SDSS sample, as well as a projected data point representing 300 SNe below redshift z=0.1. For each low-redshift sample, we assume a 0.01 systematic uncertainty

| Redshift | RMS | RMS/$\sqrt{N}$ |
|---|---|---|
| 0.0-0.1 | 0.17 | 0.0350 |
| 0.1-0.2 | 0.15 | 0.0140 |
| 0.2-0.3 | 0.14 | 0.0082 |
| 0.3-0.4 | 0.16 | 0.0073 |
| 0.4-0.5 | 0.17 | 0.0068 |
| 0.5-0.6 | 0.18 | 0.0075 |
| 0.6-0.7 | 0.18 | 0.0086 |
| 0.7-0.8 | 0.21 | 0.0120 |
| 0.8-0.9 | 0.23 | 0.0150 |
| 0.9-1.0 | 0.25 | 0.0180 |
| 1.0-1.1 | 0.21 | 0.0200 |
| 1.1-1.2 | 0.17 | 0.0240 |

Table 14:: The Hubble diagram RMS scatter, and RMS/$\sqrt{N}$, for the simulated DES hybrid 10-field survey.

in the absolute SNIa brightness (see Appendix B).

### 8.1. Figure of Merit

Constraints on cosmological parameters are obtained by comparing the theoretical values of distance moduli, $\mu(z, \theta_c)$, to the values inferred from the light curve fits of the SN simulations, $\mu_{fit}(z)$, where: $\theta_c \equiv \{\Omega_{DE}, w_0, w_a, \Omega_k\}$ is the set of cosmological parameters. The likelihood for an individual SN at redshift $z_i$, $L(\mu_{fit}|z_i, \theta_c)$, is taken to be Gaussian with a mean given by the $\mu(z_i, \theta_c)$ at redshift $z_i$, for the cosmological parameters $\theta_c$, with a standard deviation $\sigma_i^\mu$ given by the MLCS2k2 light curve fit errors and an intrinsic dispersion $\sigma_{int} = 0.13$ added in quadrature. In the case of SNe with photometrically determined redshifts, we add an error of $|\frac{\partial \mu(z, \theta_c)}{\partial z} \delta z|$ in quadrature. The simulated SN observations are independent, and the likelihood is analytically marginalized over the nuisance parameter combination of the Hubble Constant, $H_0$, and the absolute magnitude, $M$, with a flat prior. This results in a likelihood for the $\mu_{obs}(z)$ for all SNe that is Gaussian and has a covariance matrix $Cov$, which can be calculated from the above errors $\sigma_i^\mu$ on each SN discussed above.

Following the Dark Energy Task Force (DETF) Report (Albrecht et al. 2006), we evaluated the performance of survey options in terms of the DETF Figure of Merit (FoM), albeit with the following modification. The DETF FoM is de-fined to be the inverse of the area of the 95% confidence-level error ellipse in the $w_0, w_a$ plane when all other parameters have been marginalized over. However, the FoM is often calculated as $[\sigma_{w_p}\sigma_{w_a}]^{-1}$, or equivalently $[\det(\text{Cov}_{w_0 w_a})]^{-0.5}$. We follow the latter convention, as also used by SNLS (Sullivan et al. 2011). Our FoMs may be converted to the DETF FoM defined by Albrecht et al. (2006) upon dividing by factor of 18.8. Henceforth, we refer to our modified FoM as the DETF FoM. Along with DES SNe, and the low-redshift samples discussed above, we used prior constraints from the DETF Stage II experiments plus Planck in the form a Fisher Matrix for these experiments obtained from the DETF (Wayne Hu, private communication). Henceforth, we refer to this prior as "Stage II." It should be noted that this prior constrains cosmological parameters much more strongly than current data. Thus, priors used in parameter estimates based on current data releases, such the SNLS results (Sullivan et al. 2011), are much weaker, and consequently the FoMs computed from such surveys are much lower due to the weaker priors. The error covariance matrix $C$ on all the parameters is estimated as the inverse of the Fisher matrix $F$ evaluated at a fiducial set of parameters $\Theta_p$ (which is chosen to be the set suggested by DETF, i.e., $\Omega_{DE} = 0.73, \Omega_K = 0, w_0 = -1, w_a = 0$):

$$\begin{aligned} F_{ij}(\Theta_p) &\equiv F_{ij}^{DES} + F_{Stage\ II} \\ F_{ij}^{DES}(\Theta_p) &\equiv \left\langle -\partial_i \partial_j \ln(L^{DES}(\mu_{obs}|\theta_c))\Big|_{\Theta_p} \right\rangle \\ &= \frac{\partial \mu^a}{\partial \Theta_c^i} Cov_{ab}^{-1} \frac{\partial \mu^b}{\partial \Theta_c^j}\Big|_{\Theta_p} + \\ &\quad \frac{\partial^2 \ln(\det(\text{Cov}))}{2 \partial \Theta_i \partial \Theta_j}\Big|_{\Theta_p}, \end{aligned} \qquad (3)$$

where $a, b$ index each SN and $i, j$ index the four cosmological parameters. The calculated DETF FoM, assuming spectroscopic host-galaxy redshifts and statistical uncertainties only[19], ranges from 214 to 228 (see Tab. 15). The Stage II experiments plus Planck, without any additional data, yield a FoM of 58. Hence, with statistical uncertainties only, the relative improvement is by a factor of 3.69 to 3.93. In our FoM estimates, we include a reduction in the SN sample size due to incompleteness in the sample of host galaxies based on the fractions in Tab. 7. The FoM before trimming is typically a factor of 1.07 larger. Augmenting the sample with photometric redshifts, described in §4.3, results in an increase in relative FoM by a factor of 1.03. The hybrid 5-field is presented for both `MLCS2k2` and `SALT2` simulations and analyses. The difference in FoM between the two is due to the smaller Hubble residuals for the `MLCS2k2` model with a dust extinction prior (as shown in Fig. 16). In the next section, we augment our FoM calculations via the inclusion of systematic uncertainties, both with and without the effect of the dust prior.

| DES SNIa Data Set | DETF FoM (Stats.) |
|---|---|
| Hybrid 10-field | 228 |
| Hybrid 5-field | 225 |
| Hybrid 5-field (`SALT2`) | 200 |
| Shallow 9-field | 218 |
| Deep 3-field | 214 |

Table 15:: DETF Figure of Merit (modified as described in this section) for four of the DES SN survey strategies considered (see Tab. 3) using statistical uncertainties only. The results are for the `MLCS2k2` model unless otherwise noted, and include the assumed DETF Stage II plus Planck combined Fisher matrix. SN statistics are given in the Fig. 10 caption. Each survey is augmented by a projected low-redshift SNIa anchor and a simulated 3-year SDSS SNIa data set. The number of SNIa in each survey is also reduced due to host galaxy sample incompleteness based on the fractions in table 7. The Figure of Merit before trimming is typically 15 units larger.

---

[19] As explained in Appendix B, the calculation includes marginalization over the absolute magnitude.

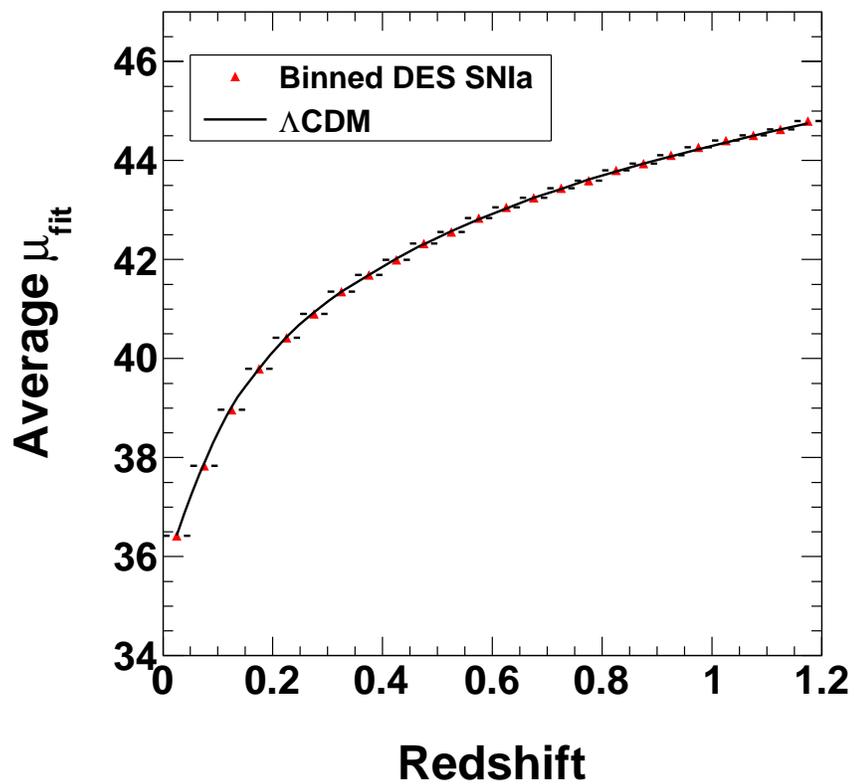

Fig. 25.—: Hubble diagram of binned SNIa for the hybrid 10-field survey. Note that the scale of the errors on the points is not visible. The RMS and RMS/$\sqrt{N}$ values for each redshift bin are shown in Tab. 14.

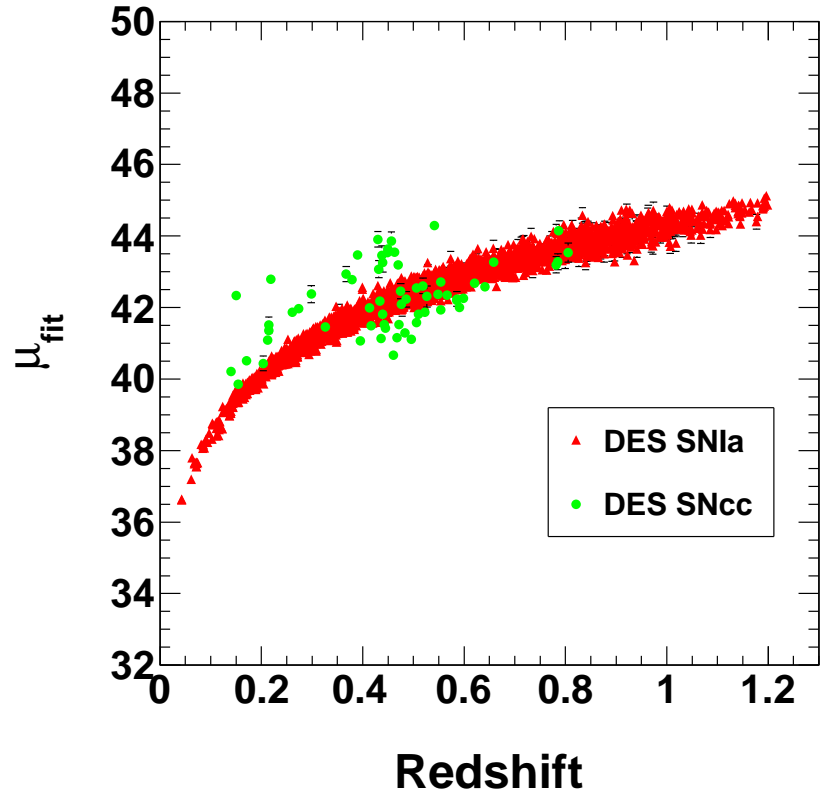

Fig. 26.—: Hubble diagram of individual SNIa and SNcc for the hybrid 10-field survey.

### 8.2. Systematic uncertainties

The DES SN search precision will depend strongly on our ability to control systematic uncertainties. In this section, we discuss the inclusion of such uncertainties in our FoM forecasts (the details of the calculation are given in Appendix B). Unless specified, the numbers in this section assume an accurate redshift derived from a host-galaxy spectrum. We consider three other fundamental sources of systematic uncertainties and one tied to an analysis option in this paper:

- filter zeropoints (fundamental);
- filter centroid wavelength shifts (fundamental);
- core collapse SNe in the SNIa sample (fundamental);
- the use of a dust prior for $R_V$ and $A_V$ (derived from an analysis choice).

In addition, all of the calculations in this section include the inter-calibration of the low-redshift anchor data sets and DES. We acknowledge the existence of astrophysical systematic effects that are not included here. Such effects are community-wide concerns and are beyond the scope of this study.

The filter zeropoint uncertainties are taken as independent and to have the value of 0.01 magnitudes (mags), which is an estimate of the final survey precision. The shift in the distance modulus is computed for each filter zeropoint change. The effect of a change in the $i$-band filter, and the corresponding change in $\mu$, is displayed in Fig. 27. Two sets of data points are shown, one for a 0.01 mag shift in $i$-band, and the other a 0.1 mag shift but with the $\mu$ change divided by 10. This demonstrates the linearity in the $\mu$ change for a shift in zeropoint. A Markov Chain Monte Carlo cosmology calculation[20], with four independent cosmology parameters, was used to evaluate the shift in the maximum likelihood value of $w_0$ due to a shift in filter zeropoint (see Fig. 28). These cosmology shifts are meant as an example and are not used in the FoM calculation described earlier and in Appendix B. The FoM including the $\mu$ changes caused by filter zeropoint shifts, as illustrated in Fig. 27, is shown in Tab. 16 for the hybrid 10-field survey. This is the most important of the fundamental systematic uncertainties listed above.

The systematic effects have also been evaluated for the hybrid 5-field survey, and the impact on the FoM was found to be very similar to that for the 10-field survey. Thus, from the point of view of constraining the CPL parameters, after combining with prior data, the two strategies are essentially equivalent. The strategy choice is then motivated by the potential for other kinds of studies that can, e.g., test and verify the accuracy of SNIa light curve models and of the redshift independence of SNIa standardized luminosities. Such studies typically require a large sample size, so that one can study correlations with other observables, e.g., SNIa host properties. The 10-field survey provides a much larger number of well-measured SNIa at redshift ranges where the potential for observing host properties, or obtaining SNIa spectra, is high. Thus, the 10-field hybrid strategy is more suitable for such studies.

| Systematic change included | FoM with systematic |
|---|---|
| None | 228 |
| Filter zeropoint shift | 157 |
| Inter-calibration | 188 |
| Filter $\lambda$ shift | 179 |
| Core collapse misid. | 226 |
| $R_V$ and $\tau_{A_V}$ | 128 |
| Total without $R_V$ and $\tau_{A_V}$ | 124 |
| Total with $R_V$ and $\tau_{A_V}$ | 101 |

Table 16:: DETF Figure of Merit (modified as described in §8.1) for the MLCS2k2 model including various systematic changes in the DES SNIa hybrid 10-field survey (including a low-redshift anchor and a simulated SDSS sample). The 5-field hybrid total values without and with $R_V$ and $\tau_{A_V}$ are 120 and 94, respectively.

The systematic uncertainties in the filter centroids are derived in a similar fashion to the zeropoints, using 10 angstroms as the expected wavelength precision for the DES. The resulting FoMs are also presented in Tab. 16.

---

[20]SNCOSMO is available as part of the SNANA package.

The systematic uncertainty due to SNcc misidentification is caused by the fitted-$\mu$ difference between SNIa and SNcc (see Fig. 26). SNcc are generally dimmer than SNIa, and, in a fit for SNIa parameters, this causes a shift in $\mu$ to larger values. In this analysis, the fraction of SNcc in the SNIa samples is small, typically <5%. The resulting small average $\mu$ shift, and the fact that the SNcc that pass selection cuts are all at low redshift where the low-redshift anchor suppresses their effect, causes a relatively small decrease in FoM (see Tab. 16).

$\sim$0.38 of the SDSS reference value, and $\tau_{A_V}$ within $\sim$0.06 of the SDSS reference value, were used to derive the FoM. The resulting effect on the FoM is actually more significant than that of the fundamental systematics (see Tab. 16). These uncertainties can be improved by using all the color information available; on the other hand, they ignore possible redshift dependence. Our analysis indicates that the effect of the current dust prior systematic is much larger than the effect of increased Hubble residuals in the SALT2 analysis FoM (see Tab. 15).

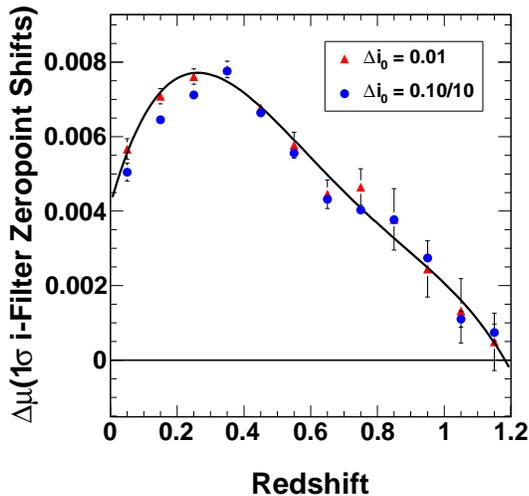

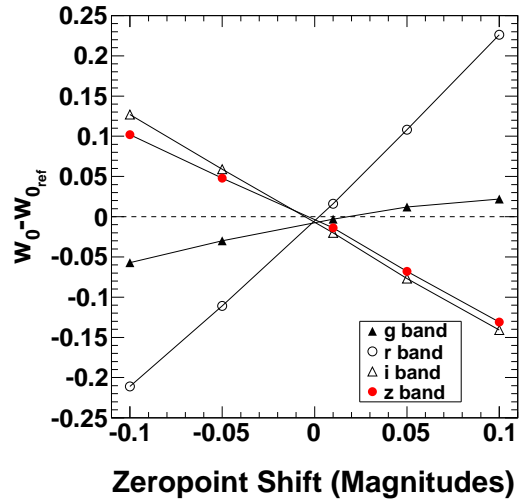

Fig. 27.—: Example shift in the average distance modulus, binned in redshift, for a 0.01 magnitude error in the $i$-band filter zeropoint assuming the DES 5-field hybrid strategy. The black line is a polynomial fit to the triangles. Also shown are the shifts (divided by 10) for a 0.1 magnitude error, demonstrating linearity in the $\mu$ change. The shift in the 10-field survey is very similar to this.

Fig. 28.—: Example shifts in $w_0$ for systematic changes in filter zeropoints assuming the DES 5-field hybrid strategy.

The total FoM, including our current estimates of systematic uncertainties, is shown in Tab. 16, for the cases both with and without the dust prior. The 95% CL limits on $w_0$ and $w_a$ are displayed in Fig. 29, for statistical uncertainties and including all systematic uncertainties. Fig. 30 displays the total 95% CL limits on $w$ as a function of redshift, and includes curves for the DETF Stage II prior alone.

Overall, the DES SNIa sample, augmented by a low-redshift anchor set, is expected to constrain a time-dependent parametrization of $w$, and improve the DETF FoM by at least a factor of 1.75 over the Stage II value of 58.

The final systematic uncertainty considered is the use of an incorrect dust extinction prior in the SNIa fitting procedure. Fig. 16 showed that the use of the prior in the MLCS2k2 fit improved the RMS scatter of the Hubble diagram, compared to the SALT2 fit which had no prior. However, the trade off is an additional systematic uncertainty since an incorrect prior in the fit can bias the distance modulus. The uncertainty in $R_V$ and $\tau_{A_V}$ is derived from the analysis of one SNIa color presented in §7.1 and in Fig. 22. Values of $R_V$ within

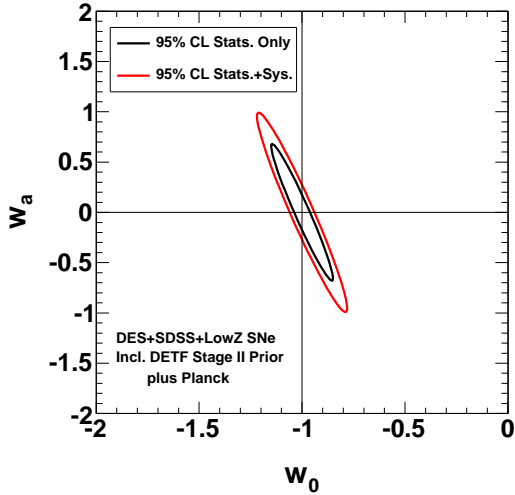

Fig. 29.—: Projected 95% limits on $w_0$ and $w_a$ as a function of redshift, with and without systematic uncertainties, assuming the DES 10-field hybrid strategy. These constraints are marginalized over $\Omega_{DE}$, $\Omega_k$, and, in the systematics case, the systematics nuisance parameters. The corresponding figure for the 5-field survey is very similar.

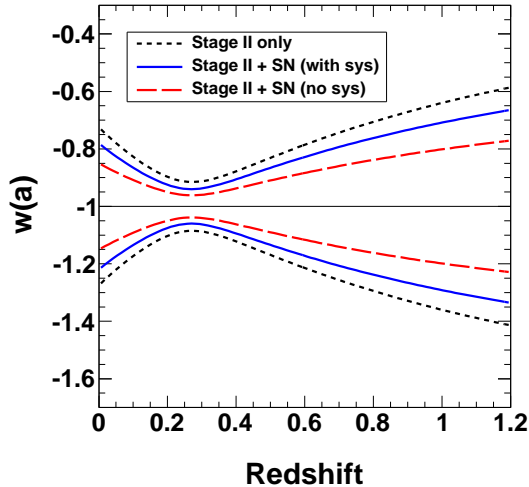

Fig. 30.—: Projected 95% limits on $w(a)$ as a function of redshift, with and without systematic uncertainties, assuming the DES 10-field hybrid strategy. Also shown is the DETF Stage II prior by itself. The corresponding figure for the 5-field survey is very similar.

## 9. Discussion and summary

We have presented an analysis of supernova light curves simulated for the upcoming Dark Energy Survey (DES) using the public SNANA package (Kessler et al. 2009b). The DES collaboration expects first light to occur in 2012. We have discussed, in detail, a prescription for the selection of a supernova search strategy prior to the onset of survey operations. We have taken several facets of observational supernova methodology into consideration, e.g., filter selection, observing field selection, cadence, exposure time, bias mitigation, sample purification, and spectroscopic and photometric redshift determination. In our analysis, we have additionally included the effects of the DES site weather history, relative position of the Moon, and observing gaps introduced by community use of the DES Dark Energy Camera. We showed that the choice of the MLCS2k2 or SALT2 supernova light curve model impacted our simulation results as follows. SALT2 simulations exhibited significantly less high-redshift bias in the distance modulus residuals than did those based on MLCS2k2 with a flat prior. However, under the assumption of the application of the correct MLCS2k2 prior, the use of SALT2 resulted in a ∼10% reduction in the statistics-only DETF Figure of Merit due to a 50%–100% increase in the RMS scatter of the high-redshift distance modulus residuals.

We forecast that the DES will discover up to ∼4000 well-measured Type Ia supernovae out to a redshift of up to 1.2, with four-pass-band photometry up to a redshift of z∼0.7–0.8. Spectroscopic redshift determination from a maximally complete host galaxy follow-up program is planned. Based on our detailed simulations, we have determined that, prior to the completion of the follow-up campaign, DES photometric redshifts will be sufficient for determining interim cosmological constraints. In addition, our projection of the ability of DES to distinguish non-Type Ia supernovae within the larger sample will lead to a Type Ia sample purity of 98% for a Type Ia fit probability cut of 0.1 assuming the 10-field hybrid survey strategy (see Tab. 3).

We have further presented two initial studies of DES supernova colors and dust extinction, as follows. First, we harnessed `SNANA` to explore the DES sensitivity to the `MLCS2k2` model parameters $A_V$ and $R_V$ based on an analysis of supernova color variations for a grid of $R_V$ and $\tau_{A_V}$. We found, for example, that the difference between the bluest and reddest filter magnitudes varies on the order of a few tenths over the commonly accepted $R_V$ range of 1.1 to 3.1. Second, we discussed the planned overlap of the DES supernova search with that of the VIDEO Survey. In order to evaluate this opportunity to obtain a combined optical+infrared supernova sample, we extended the capability of `SNANA` to simulate and fit near-infrared light curves. We found that the DES and VIDEO Survey supernova searches will yield on the order of 100 joint light curves over the anticipated two years of operational overlap.

During the course of our DES supernova strategy study, we considered a range of filter choices and survey areas. We considered use of the DES $Y$ filter for the supernova search, and found that it came at too high of a cost in terms of exposure time. This fact, coupled with the expected $YJ$ coverage from the VIDEO Survey, ultimately led us to settle on a *griz* DES supernova search. We have evaluated a suite of possible supernova survey areas from 1 to 10 DES fields (3–30 deg$^2$). This suite included a range of survey depths, within the constraint of the total supernova exposure time allocated within the larger DES, from narrow and deep, to wide and shallow, and a hybrid approach with both deep and shallow fields. For the shallow and 5-field hybrid strategy, a shallow field is defined to be one with one third of the exposure time of a deep field. Note that the depth is determined by the exposure time, as all DES supernova fields have the same 3 deg$^2$ area. Broadly speaking, the trade off between the deep and shallow strategies can be summarized in terms of the high photon statistics and redshift depth of deeper strategies and the overall number of supernovae measured for a shallow strategy. In addition, we found that deep surveys, as expected, are less susceptible to supernova selection biases than are wide surveys. In order to take advantage of the benefits of both types of survey, we have identified that a hybrid survey is a good choice for the DES.

We further dissected our supernova survey choice into the specifics of the hybrid strategy to pursue. The hybrid strategy initially considered calls for 5 fields: 2 deep field and 3 shallow fields. While this strategy offers clear benefits over the 9-field, shallow-only option considered, the question arose of how many shallow fields the hybrid should have. In addressing this question for the DES, we found that covering more area increases the number of supernovae faster than increasing the exposure time, and that the DETF FoM should be maximized by having the sample contain the maximum number of supernovae at the lowest possible redshift. These two points argue for a larger number of shallow fields. A counter to this argument arises from the fact the total amount of time allocated to the DES supernova search is fixed. This means that, for every additional shallow field, on average there is less exposure time available for each shallow field.

Our analysis motivates that having 8 shallow fields (created by dividing 1 deep field into the shallow fields with 1/8 of the exposure time) offers an attractive balance of these considerations, resulting in a 10-field hybrid strategy. Quantitatively, we have found that while the 5- and 10-field hybrid strategies yield similar DETF Figure of Merits, the primary advantages of the 10-field hybrid, as compared to the 5-field hybrid, are an increase in supernova statistics, mostly at medium DES supernova redshifts, by greater than 1/3, and a ∼75% decrease in non-Ia supernovae passing selection cuts. The first advantage of the 10-field hybrid is of key importance because supernovae at redshifts between z=0.4 and z=0.8 could form a DES supernova "calibration" sample. Our goal with this sample is to simultaneously obtain high quality follow-up spectroscopic data for both the supernovae and host galaxies. We target this redshift range for high supernova statistics because it offers an increase in redshift coverage relative to SDSS and improved $z$-band coverage compared to SNLS, thereby enhancing the competitiveness of the DES supernova sample for follow-up with leading, 8-meter-class telescopes. While pushing the DES supernova coverage to high redshift is attractive, it is expensive in terms of 8-meter-class telescope time. Such expense is not easily justified given the relatively low DES supernova signal-to-noise at high redshifts.

In closing, we forecast that the Dark Energy Survey supernova search will yield as many as 4000 well-measured Type Ia supernovae out to a red-shift as high as 1.2, with a sample purity of up to 98%. This sample will be the largest cohesive set of Type Ia supernova photometric data to date. Based on the results of the analysis in this paper, we project that the Dark Energy Survey supernova search will attain DETF Stage III status (Albrecht et al. 2006) by improving the DETF Figure of Merit by a factor of at least 1.75 relative to DETF Stage II experiments.

We thank John Cunningham of the Loyola University Chicago Department of Physics for his assistance with the completion of the simulations that form the foundation of this paper. In addition, we thank Delphine Hardin for supplying the SNLS information included in Tab. 19. We also thank the anonymous referee for insightful comments that led to significant improvement of the paper.

The submitted manuscript has been created by UChicago Argonne, LLC, Operator of Argonne National Laboratory ("Argonne"). Argonne, a U.S. Department of Energy Office of Science laboratory, is operated under Contract No. DE-AC02-06CH11357. The U.S. Government retains for itself, and others acting on its behalf, a paid-up nonexclusive, irrevocable worldwide license in said article to reproduce, prepare derivative works, distribute copies to the public, and perform publicly and display publicly, by or on behalf of the Government.

This paper has gone through internal review by the DES collaboration. Funding for the DES Projects has been provided by the U.S. Department of Energy, the U.S. National Science Foundation, the Ministry of Science and Education of Spain, the Science and Technology Facilities Council of the United Kingdom, the Higher Education Funding Council for England, the National Center for Supercomputing Applications at the University of Illinois at Urbana-Champaign, the Kavli Institute of Cosmological Physics at the University of Chicago, Financiadora de Estudos e Projetos, Fundação Carlos Chagas Filho de Amparo à Pesquisa do Estado do Rio de Janeiro, Conselho Nacional de Desenvolvimento Científico e Tecnológico and the Ministério da Ciência e Tecnologia, the Deutsche Forschungsgemeinschaft and the Collaborating Institutions in the Dark Energy Survey.

The Collaborating Institutions are Argonne National Laboratories, the University of California at Santa Cruz, the University of Cambridge, Centro de Investigaciones Energeticas, Medioambientales y Tecnologicas-Madrid, the University of Chicago, University College London, DES-Brazil, Fermilab, the University of Edinburgh, the University of Illinois at Urbana-Champaign, the Institut de Ciencies de l'Espai (IEEC/CSIC), the Institut de Fisica d'Altes Energies, the Lawrence Berkeley National Laboratory, the Ludwig-Maximilians Universitt and the associated Excellence Cluster Universe, the University of Michigan, the National Optical Astronomy Observatory, the University of Nottingham, the Ohio State University, the University of Pennsylvania, the University of Portsmouth, SLAC, Stanford University, the University of Sussex, and Texas A&M University.

## A. Fractions of SNIa Host Galaxies Satisfying Apparent-Magnitude Limits

In order to determine the spectroscopic follow-up requirements for the DES SN search, it is necessary to know, as a function of redshift, the numbers of SNIa host galaxies that satisfy the $i$-band apparent-magnitude limits discussed in §4. In general, it is useful to quote the fractions of SNIa host galaxies in a given redshift interval that fall into the various classes. For any given survey, these fractions can then be combined with the expected redshift distribution of SNIa to obtain, for each redshift interval, the expected number of SNe that will require spectroscopic follow-up. The galaxy fractions are calculated from the luminosity distributions of SNIa host galaxies. These luminosity distributions can be estimated by weighting the luminosity distributions of field galaxies by the probability that a galaxy will host a SNIa. We assume that the luminosity of a host galaxy scales with its stellar mass, we make the ansatz that this probability is proportional to the host-galaxy stellar-mass dependence of the measured rate of SNIa.

We have made a number of simplifying assumptions in the calculation presented here. First, we have assumed the presence of sharp cut-offs in the apparent-magnitude limits that determine whether or not a galaxy will have a measured follow-up spectrum or a photo-$z$ estimate. In reality, this will not be the case. Due to various inefficiencies, spectra will not be obtained for all galaxies with $m_i < 24$. For example, the DEEP2 survey (Faber et al. 2007) quote an overall efficiency for obtaining follow-up spectra of about 70%. On the other hand, spectra will be obtained for some galaxies that are dimmer than $m_i=24$, but have strong emission lines. We did not include the misidentification of a SN host galaxy, both because it is small (Smith et al. 2011, find this to be a 2% effect at low redshift), and because our analysis is rather insensitive to this effect. The latter point is due to the precision of DES photometric redshifts, which is of particular importance at high redshifts where host identification uncertainties are the largest. A second assumption in our calculation is that we can ignore variations in the surface-brightness of galaxies and that the parameters that we use to characterize the behavior of the luminosity functions are free from surface-brightness selection effects. Third, in determining the probability that a galaxy can host a SNIa, we assume that we can ignore the star-formation rate. Instead, we assume that we can derive a probability based solely on the stellar-mass dependence of the SNIa rate. This assumption introduces uncertainties which we estimate by choosing a range in the mass dependence that covers the measured values for star-forming and passive galaxies. Finally, we make some simplifying assumptions in the treatment of K-corrections. For galaxies with measured follow-up spectra, the K-corrections are irrelevant, since the entire spectrum will be measured. The apparent-magnitude limits that we are using to estimate the fractions are simply a convenient way to quantify the capabilities of the telescopes that are used to obtain the follow-up spectra. However, for a host galaxy with a photometrically determined redshift, K-corrections can significantly reduce its $i$-band apparent magnitude, and hence can impact the precision of its measured photo-$z$. K-corrections vary significantly depending on the galaxy morphology, but typically become more important around a z of 0.7, where the 4000 $A°$ break in the galaxy SED crosses into the $i$-band. A full treatment of K-corrections is beyond the scope of this study. Below, we give some estimates of the effect for different galaxy types. These estimates are based on the assumption that we can use a simple linear approximation to characterize the typical shape of the SED for each galaxy type. With these caveats in mind, we now present the details of our calculation.

Field galaxies have luminosity density functions that are well described by Schechter functions of the form

$$\phi(M)dM = 0.4\log(10)\phi^* 10^{0.4(M^*-M)(\alpha^*+1)} \exp(-10^{0.4(M^*-M)})dM, \tag{A1}$$

where $M$ is the absolute magnitude of the host galaxy in some filter, and $\phi^*$, $M^*$, and $\alpha^*$ are experimentally measured parameters. These parameters are usually quoted for rest-frame filters. However, since we are interested in DES $i$-band magnitude limits, it is convenient to use the $i$-band Schechter-function parameters from Blanton et al. (2003) because the wavelength range for the $i$-band filter in SDSS is very close to that for DES. For now, we consider the case where the parameters are fixed. Below, we will address the case where they evolve with $z$. The dependence of the rate of SNIa on the host-galaxy stellar mass, $M_\star$, is usually parametrized as a power law of the form $M_\star^{\kappa_{Ia}}$. A summary of the measured values of $\kappa_{Ia}$ for different host-galaxy types is given in Tab. 17.

| Reference | Host Galaxy Type | $\kappa_{Ia}$ |
|---|---|---|
| Sullivan et al. (2006) | Passive | $1.10 \pm 0.12$ |
| Sullivan et al. (2006) | Strongly Star Forming | $0.74 \pm 0.08$ |
| Sullivan et al. (2006) | Weakly Star Forming | $0.66 \pm 0.08$ |
| Smith et al. (2011) | Passive | $0.67 \pm 0.15$ |
| Smith et al. (2011) | Star-Forming | $0.94 \pm 0.08$ |
| Li et al. (2011b) | All | 0.5 |

Table 17:: References for the dependence of the SNIa rate on the host-galaxy stellar mass.

The range of values in Tab. 17 exceeds the quoted uncertainties. Furthermore, Smith et al. (2011) and Sullivan et al. (2006) disagree on the trend in values for a given type of host galaxy. We choose therefore to take the lowest and highest values from Tab. 17 as the plausible range for the stellar-mass dependence of SNIa hosts, and present fractions corresponding to this range. Assuming that the luminosity is proportional to the stellar mass, we find that the luminosity density function for SNIa host galaxies is given by

$$\phi_{Ia}(M)dM = 0.4 \log(10) \phi^*_{Ia} 10^{0.4(M^* - M)(\alpha^* + \kappa_{Ia} + 1)} \exp(-10^{0.4(M^* - M)}) dM, \tag{A2}$$

where $\phi^*_{Ia}$ is an (unknown) normalization constant that is assumed to be proportional to $\phi^*$, and $\kappa_{Ia} = 0.5$ or 1.10. Eqn. A2 predicts an absolute-magnitude distribution of SNIa host galaxies that is in qualitative agreement with the distribution measured by Yasuda & Fukugita (2010).

The next step is to determine, in the presence of an apparent magnitude cut, $m_{lim}$, the number of SNIa host galaxies that will be seen in a thin shell at redshift $z$. All galaxies having absolute magnitudes brighter than $M = m_{lim} - \mu(z) - K(z)$ will be visible. Here, $\mu(z)$ is the distance modulus that is determined from the redshift assuming a particular cosmology, and $K(z)$ is the K-correction that accounts for the redshift of the galaxy spectra. Hence the fraction of visible galaxies is given by

$$\frac{\int_{-\infty}^{m_{lim} - \mu(z) - K(z)} \phi_{Ia}(M) dM}{\int_{-\infty}^{\infty} \phi_{Ia}(M) dM}. \tag{A3}$$

We note that the normalization constant, $\phi^*_{Ia}$, in Eqn. A2 cancels in this fraction. Furthermore, for the values of $\alpha^*$ that are typically measured from field-galaxy data, the integrals in Eqn. A3 are convergent for large $M$ only because of the extra terms in the integrands that are dependent on $\kappa_{Ia}$. Integrating Eqn. A3 over $M$ and $z$ yields the fraction of visible galaxies in the range $z_{lo} < z < z_{hi}$.

$$f(z_{lo}, z_{hi}, m_{lim}) = \frac{\int_{z_{lo}}^{z_{hi}} \Gamma(\alpha^* + \kappa_{Ia} + 1, 10^{0.4(M^* - m_{lim} - \mu(z) - K(z))}) dV_{co}}{\Gamma(\alpha^* + \kappa_{Ia} + 1) \int_{z_{lo}}^{z_{hi}} dV_{co}}, \tag{A4}$$

where $\Gamma(s, x)$ and $\Gamma(s)$ are the upper incomplete Gamma function and the Gamma function, respectively, and $dV_{co}$ is the co-moving volume element.

Given a cosmology, Eqn. A4 can now be evaluated over any desired redshift range to give the SNIa host fractions. In Tab. 18, we present these fractions for a $\Lambda$CDM cosmology with $\Omega_\Lambda = 0.73$, $\Omega_m = 0.27$, $H_0 = 72$ km s$^{-1}$ Mpc$^{-1}$. We choose redshift intervals of 0.1 and the limiting cases of $\kappa_{Ia}$=0.5 and $\kappa_{Ia}$=1.1. We set $K(z) = 0$ for now. Two regions of the host-galaxy i-band apparent magnitude, $m_i$, are considered: $m_i < 24$, $24 < m_i < 26$. These regions correspond to the current expectations for the apparent magnitude limits for which DES will be able to obtain spectroscopic and photo-z host-galaxy redshifts, respectively. SNIa whose host galaxies do not fall into either of these two classes will either have no visible hosts, or have hosts with large uncertainties in their photo-z redshifts.

So far, we have assumed that the parameters characterizing the shape of the Schechter functions in Eqns. A1 and A2 do not vary with redshift. In fact, current measurements indicate that the parameters $\phi^*$, $M^*$, and $\alpha^*$ all vary with $z$.(Blanton et al. 2003; Faber et al. 2007; Li et al. 2011b; Poli et al. 2003). If the normalization, $\phi^*$ is dependent on $z$, it no longer cancels exactly in Eqn. A4. However, we have checked that the change in the fractions is much less than 1% for the values of $\phi^*(z)$ that have been measured from existing data. Hence we can safely ignore any changes in $\phi^*$. Variations in $\alpha^*$ affect the shape of the Schechter function at large values of $M$. Most studies have kept the value of $\alpha^*$ fixed. The limited data that are available for $\alpha^*$ evolution (Poli et al. 2003) show a modest decrease in the value of alpha for $B$-band measurements. Since we do not have any $i$-band measurements of $\alpha^*$ evolution and, as can be seen from Eqn. A4, changes in the value of $\kappa_{Ia}$ have the same effect as changes in $\alpha^*$, for the purposes of this analysis, we can ignore evolution in $\alpha^*$. Evolution of $M^*$ can be parametrized as $M^*(z) = M^*(z_0) - Q(z - z_0)$, where $z_0$ is some reference redshift. This parametrization is very convenient because Eqn. A4 is still correct, once $M^*$ is replaced by $M^*(z)$. The measured values of $Q$ are filter dependent, have large uncertainties and show a substantial variation with galaxy type. Lin et al. (1999), in the CNOC2 survey, found that Early-type galaxies have larger, positive values of $Q$ of $O(1-2)$, whereas Late-type galaxies have smaller values of $Q$ less than 0.5. Note that since $Q$ is positive, galaxies get brighter as $z$ gets larger, so ignoring the effects of evolution will lead to overestimates of the number of galaxies that would fail any apparent magnitude cut. Blanton et al. (2003), in the SDSS survey, use a more complicated parametrization for the evolution of the galaxy luminosity-density functions. However, their parametrization is reasonably well approximated by a simpler Schechter-function parametrization at their reference redshift of $z_0=0.1$. Since we assume that this will also be the case at higher redshifts, we can still use the value of 1.6 that they fit for their $Q$-parameter as an estimate for the $Q$-value in our linear-evolution case. We therefore show the effects of $z$-evolution on the SNIa host-galaxy fractions by choosing values of $Q$ equal to 0, 0.5 and 1.6, which are representative of the range of values present in the data. In Table 19, we present the fractions for these three values of $Q$ and $\kappa_{Ia} = 0.74$. Following Blanton et al. (2003), we choose $z_0 = 0.1$.

| $z$-range | $\kappa_{Ia} = 0.5$ | | $\kappa_{Ia} = 1.10$ | |
|---|---|---|---|---|
| | $m_i < 24$ | $24 < m_i < 26$ | $m_i < 24$ | $24 < m_i < 26$ |
| $0.1 - 0.2$ | 0.93 | 0.04 | 0.998 | 0.002 |
| $0.2 - 0.3$ | 0.89 | 0.07 | 0.994 | 0.005 |
| $0.3 - 0.4$ | 0.84 | 0.10 | 0.986 | 0.012 |
| $0.4 - 0.5$ | 0.78 | 0.13 | 0.975 | 0.022 |
| $0.5 - 0.6$ | 0.73 | 0.16 | 0.958 | 0.037 |
| $0.6 - 0.7$ | 0.67 | 0.19 | 0.935 | 0.056 |
| $0.7 - 0.8$ | 0.61 | 0.23 | 0.906 | 0.081 |
| $0.8 - 0.9$ | 0.55 | 0.26 | 0.87 | 0.11 |
| $0.9 - 1.0$ | 0.50 | 0.29 | 0.83 | 0.14 |
| $1.0 - 1.1$ | 0.44 | 0.32 | 0.78 | 0.18 |
| $1.1 - 1.2$ | 0.39 | 0.34 | 0.73 | 0.23 |

Table 18:: Fractions of the total number of SNIa host galaxies for various apparent magnitude limits and values of $\kappa_{Ia}$.

As mentioned above, K-corrections will reduce the $i$-band apparent magnitudes of host galaxies, particularly those above a redshift of 0.7, which is where the 4000 $\text{\AA}$ break crosses into the rest-frame $i$-band. If we now include K-corrections in our estimates, the fractions with $m_i < 24$ decrease, and the fractions with $24 < m_i < 26$ increase. In general, because they have flatter SEDs, the size of the changes are larger for elliptical galaxies than for star-forming galaxies. If we assume that the SED of a strongly star-forming galaxy rises linearly by a factor of 3 from 8500 $\text{\AA}$ down to the 4000 $\text{\AA}$ break, and then falls by a factor of about 2 below the break, we find that the fractions for the middle column of Table 19 change by less than a few percent below $z = 0.7$. Above $z = 0.7$, the effect increases with $z$, and results in a 16% decrease in the $m_i < 24$ estimate and a 17% increase in the $24 < m_i < 26$ estimate for $z$ between 1.1 and 1.2. On the other hand, if we assume that the SED of an elliptical galaxy is flat and falls by a factor of 2 below the 4000 $A°$ break, we find that the effects are much larger. Even below $z=0.7$, the changes grow with $z$, up to a 10% decrease in the $m_i<24$ estimate and a 6% increase in the $24<m_i<26$ estimate for $z$ between 0.6 and 0.7. Above $z=0.7$, the effect of the corrections again rises with $z$ and results in a 40% decrease in the $m_i<24$ estimate and a 24% increase in the $24<m_i<26$ estimate for $z$ between 1.1 and 1.2. The effect of K-corrections on the galaxy fractions should lie somewhere between these two extremes, depending on the precise mix of SNIa host-galaxy morphologies. Hence, the size of this uncertainty is comparable to the other uncertainties discussed in this section. Note, however, that these uncertainties affect only the estimates of the size of the photo-$z$ sample. We have seen in §8.1, that the FOM decreases by only about 15 units, even if no SNe in the photo-$z$ sample are included. Hence, adding some fraction of SNe from the photo-$z$ sample to the analysis can result in only very modest gains in the FOM, and we expect that our forecasts for the FOM values to be largely unaffected by the uncertainties in the K-corrections.

| $z$-range | $Q=0$ | | $Q=0.5$ | | $Q=1.6$ | |
|---|---|---|---|---|---|---|
| | $m_i<24$ | $24<m_i<26$ | $m_i<24$ | $24<m_i<26$ | $m_i<24$ | $24<m_i<26$ |
| $0.1-0.2$ | 0.98 | 0.01 | 0.98 | 0.01 | 0.99 | 0.01 |
| $0.2-0.3$ | 0.96 | 0.03 | 0.97 | 0.03 | 0.97 | 0.02 |
| $0.3-0.4$ | 0.94 | 0.05 | 0.94 | 0.04 | 0.95 | 0.04 |
| $0.4-0.5$ | 0.91 | 0.07 | 0.92 | 0.06 | 0.94 | 0.05 |
| $0.5-0.6$ | 0.87 | 0.10 | 0.89 | 0.09 | 0.92 | 0.06 |
| $0.6-0.7$ | 0.82 | 0.13 | 0.85 | 0.11 | 0.90 | 0.07 |
| $0.7-0.8$ | 0.77 | 0.17 | 0.82 | 0.14 | 0.89 | 0.09 |
| $0.8-0.9$ | 0.72 | 0.20 | 0.78 | 0.16 | 0.87 | 0.10 |
| $0.9-1.0$ | 0.67 | 0.24 | 0.74 | 0.19 | 0.86 | 0.10 |
| $1.0-1.1$ | 0.61 | 0.28 | 0.70 | 0.22 | 0.85 | 0.11 |
| $1.1-1.2$ | 0.55 | 0.32 | 0.67 | 0.24 | 0.84 | 0.12 |

Table 19:: Fractions of the total number of SNIa host galaxies for various apparent magnitude limits for $\kappa_{Ia}=0.74$ and three values of $M^*$-evolution parameter, $Q$. The chosen values of $Q$ span the range of possible values found in the field-galaxy data.

We conclude that uncertainties in the data that can be used to constrain the SNIa host-galaxy fractions are large and lead to big variations in the estimates of the number of host galaxies that fall into the various magnitude classes. As discussed above, we have made several simplifying assumptions in our calculations that ignore a number of known effects due to inefficiencies in redshift determinations, galaxy morphology, galaxy surface brightness and star-formation rate. We have attempted to include the effects due to $z$-evolution of the Schechter functions, but here too there are dependencies on galaxy morphology that contribute to the uncertainties on the fractions. We quote numbers in the mid-range of the predictions in the body of the paper, but note that predictions at high $z$ vary by about 25%.

## B. SNe systematic uncertainties in the FoM calculation

The use of a Fisher Information Matrix in forecasting constraints from a survey has been discussed in several works. While the general formalism can account for certain kinds of systematic uncertainties in the forecasted constraints, the specific examples for supernova surveys that are usually studied only account for statistical uncertainties. The DETF report suggested accounting for systematic uncertainties in forecasting cosmological constraints from supernova surveys. In particular, the DETF report discussed four kinds of systematic uncertainties: 1. photo-z errors, 2. absolute magnitude uncertainties, 3. step $\mu$ offset for the near sample, 4. quadratic $\mu$ offset (see subsection below for additional information). With the advantage of our simulations and knowledge specific to the setup of DES, we extend the DETF analysis by choosing a more appropriate set of systematics particularly with respect to their fourth systematic parameter. In this appendix, we explain the choices made in our paper and how they relate to the DETF choices.

The basic idea is that the naive estimator for the distance modulus as obtained from the light curve fitter is biased when the survey setup and conditions or the properties of SNIa are actually different from those assumed in the analysis. However, since the survey conditions can be varied in our simulation inputs, we can estimate the bias $\Delta\mu(z, \Theta)$, which is the expected deviation of the estimated distance modulus from the true distance modulus, for each set of conditions. We thus correct for the bias by replacing $\mu(z, \theta_c) \to \langle(\mu, z\Theta)\rangle = \mu(z, \theta_c) - \Delta\mu(z, \Theta)$, where $\Theta$ includes not only the cosmological parameters, $\theta_c$, but also the set of systematic parameters, $\theta_n$, modeling the setup and SN properties. These systematic parameters are measured independently, and the information from these measurements may be treated as a Gaussian prior resulting in a prior Fisher matrix $F_{ind}$. Extending the set of model parameters $\Theta$ to include these systematic parameters $\theta_n$ as "nuisance parameters", we can use the replacement in computing the Fisher matrix and then marginalize over the allowed values of these nuisance parameters. We give the expression for the Fisher matrix

$$\begin{aligned}
F_{tot} &= F_{ij} + F_{Stage\ II} + F_{ind}, \\
F_{ij}(\Theta_p) &= \left.\frac{\partial\langle\mu^a(z,\Theta)\rangle}{\partial\Theta_i}\right|_{\Theta_p} (Cov^{-1})_{ab} \left.\frac{\partial\langle\mu^b(z,\Theta)\rangle}{\partial\Theta_j}\right|_{\Theta_p} + \frac{1}{2}\frac{\partial^2 \ln(\det(Cov))}{\partial\Theta_i \partial\Theta_j}
\end{aligned} \quad (B1)$$

where $a, b$ index the SN and are summed over in the above equation, and $Cov$ is the covariance matrix used in the (statistics only) Fisher matrix. The second term arises from the derivatives of the logarithm of the normalization of the Gaussian. The normalization of the Gaussian distribution only depends on cosmological parameters through the term $\frac{\partial\mu}{\partial z}\delta z$ added in quadrature to the light curve fit errors. For the cases where supernova redshifts are determined spectroscopically leading to small $\delta z$, the covariance matrix $Cov$ becomes independent of the cosmological parameters and hence the second term is zero. For cases, where the supernova redshift is determined photometrically, this turns out to be a very small contribution. All partial derivatives are evaluated at a set of fiducial parameters $\Theta_p$ taken to be the DETF parameters $\{\Omega_{DE}, w_0, w_a, \Omega_k\} = \{0.73, -1, 0, 0\}$ for the cosmological parameters, and our best estimates for the nuisance parameters. We note that this is exactly the idea behind the treatment suggested in the DETF report. We now proceed to discuss our choice for the set of nuisance parameters as described below.

### B.1. Nuisance Parameters

The main differences in our method, compared to that presented in the DETF report, are as follows:

1. We have an improved estimate of the effect of photo-z uncertainties, coming from the full simulation of the DES survey.

2. The uncertainty in absolute magnitudes referred to in the DETF report is analytically marginalized in our likelihood function. Therefore, our "statistics only" Fisher matrix accounts for this in the correlated covariance matrix.

3. In our SN analysis, we include two low-redshift samples as anchors: the 3-year `SDSS` sample with ~350 SNe and a sample of 300 supernovae taken to be at redshift of 0.055 (Li et al. 2011a), each with a Gaussian error of 0.13 magnitudes . Neither of these samples were included in the calculation of the Stage II prior Fisher matrix. In each of these low-redshift samples, the dominant systematic uncertainty is expected to be a step $\mu$ offset of the kind described in the DETF report. Therefore, we include a step offset for each of the low-redshift samples. We also assume a Gaussian prior on each of these step offsets of width $\delta M_{lowz} = \delta M_{SDSS} = 0.01$. This is consistent with the suggestion of the DETF report.

4. In the DETF report, all other systematic effects were treated as an effective linear and quadratic shift in $\mu(z)$. The relevant nuisance parameters were taken to be the first and second order redshift coefficients.

We have used a more realistic estimate of the effects of systematics on $\mu(z)$. In our simulations, we vary the systematic parameters $\theta_n$, which model the instrumental setup and supernova properties, from their assumed values and study how the $\mu^{obs}(z)$ changes. Assuming that we are in the linear regime, we write

$$\mu^{obs}(z,\Theta) = \mu^{obs}(z,\Theta_p) + \sum_j \left.\frac{\partial \mu^{obs}(z,\Theta)}{\partial \theta_n^j}\right|_{\Theta_p} \Delta\theta_n^j \quad (B2)$$

where the sum runs over $j$ which indexes the list of systematic parameters $\theta_n$.

By varying each systematic parameter under consideration, one at a time, in our simulations, we estimate the average values of the partial derivatives $\frac{\partial \mu^{obs}(z,\Theta)}{\partial \theta_n^j}$ in redshift bins of 0.1 in terms of fitting functions that involves 3 to 6 parameters. Using these, we evaluate the Fisher matrix in the parameter space $\Theta$ which includes both the cosmological parameters $\theta_c$ and the systematic parameters $\theta_n$. The parameters $\theta_n$ will be set or measured to a fiducial value with an estimated uncertainty either in the process of calibration (parameters related to survey conditions) or by other experiments (SNIa light curve model parameters or SNcc fractions). These measurements allow us to use appropriate priors on the deviation of these parameters from their fiducial values. Therefore we will marginalize over all the systematic parameters with such priors on each of them. We now proceed to discuss the relevant set of systematic parameters.

We first identify the relevant set of systematic effects and parametrize them. These are the zero-points in each filter band, the wavelength of the centroid of the filters, the fraction of SNe which are SNcc but misidentified as SNIa, and the values of $\tau_{A_V}$ and $R_V$. First, we describe these parameters, and the priors on them due to independent measurements.

**Zero-Points:** We include the deviation in zero-points in each band $\{g_0, r_0, i_0, z_0\}$ from the fiducial values as parameters. Further, we expect to be able to calibrate these independently to an error of 0.01 magnitudes. Hence we assume a Gaussian prior on each of these parameters with a width $\sigma^g_{zpt} = \sigma^r_{zpt} = \sigma^i_{zpt} = \sigma^z_{zpt} = 0.01$.

**Wavelength of Filter Centroid:** We include the deviation of the wavelength $\{\lambda_g, \lambda_r, \lambda_i, \lambda_z\}$ of the centroid of each filter from the fiducial values. Further, we expect to be able to calibrate these independently to an error of 10 angstroms. Hence we assume a Gaussian prior on each of these parameters with a width $\sigma^g_\lambda = \sigma^r_\lambda = \sigma^i_\lambda = \sigma^z_\lambda = 10$ Å.

**Core-Collapse Fraction:** The sample purity for the hybrid 10-field survey is 98%. We are taking the uncertainty in this value to be 2%, and have shown that the effect on the FoM is still smaller than the other uncertainties. To be specific, we use $f_{cc} = 0$ as a fiducial value and $\sigma_{f_{cc}} = 0.02$ as the Gaussian uncertainty in $f_{cc}$.

**CCM Dust Model Parameters:** $\mu_{fit}$ depends on the true values of the parameters $\tau_{A_V}$ and $R_V$ values. Our simulations are based on SDSS results (Kessler et al. 2009a) and, hence, we assume fiducial values of $\tau_{A_V}$=0.334 and $R_V$=2.18. Assuming that these variables are correlated in the way determined by the SDSS SN survey (Kessler et al. 2009a), we expect to have independent measurements determining these parameters as a correlated Gaussian with a covariance matrix $C$ with $C_{R_V R_V} = 0.1444, C_{\tau_{A_V},\tau_{A_V}} = 0.003136, C_{\tau_{A_V} R_V} = 0.0036176$.

As described in §2, these parameters are inputs to our simulations. Therefore, we can study changes in the fitted values of $\mu$ by changing these input parameters in SNANA. For each of the parameters that we shall describe, we compute $\frac{\partial \mu_{eff}(z,\Theta)}{\partial \Theta_i}|_{\Theta_p}$ by numerically evaluating the partial derivative of $\mu_{eff}(z,\Theta)$ as a function of redshift. For doing so, $\mu_{eff}(z,\Theta)$ is estimated as the sample mean of obtained values of the fitted distance modulus $\mu_{fit}$ in redshift bins of 0.1. Having obtained these estimated values, along their dispersions, we fit these values at discrete redshifts bin centroids to simple functions of the redshift. This allows us to evaluate the partial derivative at any redshift in the range of observation $(0, 1.2)$.

---

This 2-column preprint was prepared with the AAS LATEX macros v5.0.